%% file: TemsPaper.tex
\documentclass[conference]{IEEEtran}
\IEEEoverridecommandlockouts

\usepackage[ruled]{algorithm2e}
\usepackage{color,soul}

\SetAlFnt{\small}
\SetAlCapFnt{\small}
\SetAlCapNameFnt{\small}
\SetAlCapHSkip{0pt}
\IncMargin{-\parindent}

\usepackage{algorithmic}
\usepackage{amsfonts}
\usepackage{amsmath}
\usepackage{amsthm}

\usepackage{nth}
\usepackage{graphicx}
\usepackage{ragged2e}
\usepackage[caption = false]{subfig}

\def\BibTeX{{\rm B\kern-.05em{\sc i\kern-.025em b}\kern-.08em
    T\kern-.1667em\lower.7ex\hbox{E}\kern-.125emX}}

\begin{document}

\title{An Exhaustive Study of Using Commercial LTE Network for UAV Communication in Rural Areas }

\author{
	\IEEEauthorblockN{
	Mohammed Gharib\IEEEauthorrefmark{1},
	Shashidhar Nandadapu\IEEEauthorrefmark{1}, 
	Fatemeh Afghah\IEEEauthorrefmark{1}, 
	}
	\IEEEauthorblockA{\IEEEauthorrefmark{1}School of Informatics, Computing and Cyber Systems, Northern Arizona University, Flagstaff, AZ, USA\\ E-mail:\{mohammed.algharib,sn644, fatemeh.afghah\}@nau.edu}

}


\maketitle

\begin{abstract}
\input{abstract} 
\end{abstract}


\input{introduction}

\input{relatedWork}

\input{DataGathering}
\input{DataAnalysis}

\input{conclusion}

\nocite{*}
\bibliographystyle{IEEEtran}
\bibliography{References}


\end{document}

%% file: abstract.tex
Unmanned aerial vehicles (UAVs) have been increasingly used in a wide area of military and civilian applications such as data collection and monitoring. A reliable network for command and control, communication, and data transfer is crucial, not only for mission purposes but also for safety concerns. The already deployed cellular networks are appropriate candidates for UAV communication given the solid security and wide coverage of these networks. However, the reliability of such networks needs a comprehensive investigation. In this paper, we use the long-term evolution (LTE) network as the infrastructure for drone communication and data transfer, in a rural area. We study the communication characteristics of an LTE-connected drone during low-altitude flights, for different altitudes and UAV speeds. We show that, in such areas, the higher elevation benefits from a better signal quality and experiences a fewer number of handover processes. Higher speed flights also slightly degrade the communication performance\footnote{This material is based upon work supported by the Air Force Office of Scientific Research under award number FA9550- 20-1-0090 and the National Science Foundation under Grant Numbers CNS-2034218, CNS-2039026 and ECCS-
2030047.}.

%% file: introduction.tex
\section{Introduction}
Drones have been increasingly used in both military and civilian applications. Ease of deployment, highly dynamic 3D movement ability, low price, and the availability of drones make them find their way into various applications. While border surveillance, target tracking and strike are some military applications, traffic management, package delivery, search and rescue, disaster relief, and post-disaster imagery, and area monitoring are some civilian applications  to be named \cite{fireMonitoring1,fireMonitoring2}. Reliable communication is of paramount importance for command and control to guarantee the safety of drones, people and infrastructure. It is also crucial to guarantee  reliable data transfer in applications such as monitoring and surveillance.

Exploiting the existing cellular networks for UAV communications is a cost-effective solution to provide a reliable, wide-coverage, and secure communication for drones. However, the cellular networks are designed to cover the near-ground areas and might not be able to offer an optimal service to aerial users due to several concerns. The first concern is that the current cellular networks are managed to minimize the interference among  different cells' signals. The obstacles such as trees, houses, and buildings block the signals at a certain level, and hence the signal powers are managed based on the current on-ground physical situations. In higher elevations, there are no such obstacles and most of the antennas are located in the line-of-sight of the user. While the line-of-sight communication might have  stronger signal, it may cause considerable interference and make it infeasible to use such networks for drone communication. 

The second concern is about the coverage area in the higher elevations. The tilt angle of the cellular network antennas are such that the antennas face the ground to best cover the terrestrial users. Thus, the coverage area is investigated only for the two-dimension maps. In order to utilize cellular towers for drone communications, we need to investigate the coverage area above the ground and for different elevations, to find the three-dimensional coverage map. The third concern is  the  effect of the aerial users on the quality of the communications of the main terrestrial users. To the best of our knowledge, there is not enough information on the interference caused by aerial cellular users on the terrestrial users in different geographical distributions and urbanization.

In this paper, we present the results of field measurements on using LTE for drone communication in a rural area. We gather data for different low-altitude flight's elevations, $40$, $80$, and $120~m$, and different speeds of $30$ and $60$ kmph. In our measurements, we consider the call quality, and maximum achievable up-link and down-link throughput, altogether. 
The experimental measurements are performed using a commercial drone in rural area by attaching a smart phone to the drone. TEMS pocket \cite{tems} application is installed on the smart phone supported with a dedicated script on it to gather the required data. The script makes a voice call, download a large file, and  upload a stream of data to a server, using the Verizon commercial LTE network in its 1700 MHz band. All tests are done in a rural forest area close to Flagstaff city, Arizona, US. The collected log files are processed by TEMS discovery \cite{tems}. We then extracted two types of information, those of the serving cell signal, and those of comparing the best signals with one another including the serving cell signal and the neighboring cells' signals.   

In our analysis, we measured  the reference signal received power (RSRP), reference  signal received quality (RSRQ), reference  signal strength indicator (RSSI), signal to interference and noise ratio (SINR), up-link throughput, down-link throughput, and the number of handover processes. We show in our analysis that, in low-altitude flights over rural areas, the lower elevation results in the worst case performance, since the signal is attenuated with the obstacles and multi-path fading. We show that the higher elevation results in, in most cases, the best choice for the voice calls and small-size data communication. However, the moderate elevation, 80 m in our measurements, reaches the highest possible throughput in uploading and downloading. We further show that although the speed has a negligible impact on the signal quality, in most cases, the lower speed flights have slightly better performance. Furthermore, the moderate flight's elevation leads to the largest number of handovers, whereas in the highest elevation, we see the lowest number of handover processes. The higher speed flight also shows slightly less often handover processes for all the tested scenarios. Generally, we find that there are always a couple of signals with enough strength to keep the call live, and the interference caused by the line-of-sight signals does not have a significant effect on the serving cell signal quality, in the understudy settings.  

The rest of this paper is organized as follows. We review the literature works in Section (\ref{sec::relatedWork}). Then, we present the processes of  data collection, cleanup, and information extraction in Section (\ref{sec::dataGathering}). We  analyze the data and represent the results in Section (\ref{sec::dataAnalysis}). Finally, we conclude the paper and mention the future directions in Section (\ref{sec::conclusion}).

%% file: relatedWork.tex
\section{Related Work}
\label{sec::relatedWork}
Using the existing cellular networks for drone communications can facilitate the wide deployment of drone technology in a secure way without the need for substantial investments to establish new communication networks, however, the aerial coverage of cellular networks and the interference caused by the aerial users on terrestrial cellular users need to be thoroughly investigated. The third generation partnership project (3GPP) made a valuable effort in its release 15 \cite{3gpp15} to discover the support of enhanced long term evaluation (LTE) for aerial vehicles, and release 17 \cite{3gpp17} to support the 5G enhancement for UAVs. Van Der Bergh et al. \cite{Vanderbergh} studied the impact of interference and path loss on the connected drones via LTE network. They found that the signal strength is decreased rapidly by the increment in the altitude until the line-of-sight propagation is established. Concurrently, the signal quality is decreased because of the decrements in the SINR.  

Lin et al. \cite{ericsson} shared some of their measurements data for low altitude drone connected to commercial LTE network, gathered in Finland. They found that the already deployed LTE networks can support the low-altitude aerial communication, but the interference and mobility may cause challenges. Amorim et al. \cite{AmorimChannel} made several measurements to model the radio channel for aerial use of LTE network in Denmark. Their results showed better radio clearance as the aerial vehicle increases its elevation, in the low-latitude flights. However, they did not measure the SINR and left it for their future works.  

In \cite{amorimUrban}, authors performed measurements for aerial users of LTE network at elevations which do not exceed  40 m. They targeted an urban area with the highest building height of 15 m and compared their results to those of 3GPP \cite{3gpp15}. They found that the measured metrics for three different frequencies of 800, 1800, and 2600 MHz lead to similar results. 
Khawaja et al. \cite{Khawaja} found in their measurements that the signal strength mostly follows a two-path ray propagation model for higher altitudes. Al-Hourani et al. \cite{Hourani} provided a cellular to aerial channel model in terms of path loss and shadowing, based on the real experiments performed in a suburban environment.

Hayat et al. \cite{hayat} experimentally evaluated the LTE network to measure the signal to interference ratio and the downlink throughput in different elevations, in a suburban area. They showed that the throughput at 150 m altitude outperforms all lower elevation throughput. However, the throughput of 50 m elevation is much higher than that of 100 m. Kovacs et al. \cite{bellLabs} analyzed their measurements on aerial connectivity with an LTE network in a rural area. They characterized the radio channel behaviour in terms of downlink and uplink metrics, and estimated the gains of interference mitigation techniques. 

Marques et al. \cite{marques} performed an experimental study for using LTE as the communicating network for aerial vehicles, in rural areas. They measured the uplink and downlink throughput for different low-altitude elevations. They found that the 25 m elevation outperforms all other elevations in terms of uplink and downlink throughput. Muzaffar et al. \cite{raheeb} performed an experimental test for connecting the aerial user to a 5G network and compared its results with that of 4G network. They found that the 5G network generally outperform the 4G network. Overall, there is a significant need for exhaustive studies to show the usability of already deployed cellular networks for aerial communication. Most of the mentioned works performed valuable measurements and analysis. However, none of them presented a sound and complete study, covering all area types, different elevations, and all the metrics studied in this paper, altogether. Hence, we aim at studying this problem for more metrics and different elevations as well as different UAV speeds.

%% file: DataGathering.tex
\section{Data Collection and Processing} 
\label{sec::dataGathering}

\begin{figure*}[h!]
	\centering
	\subfloat[Overall view]{\fbox{\includegraphics[width=0.31\linewidth]{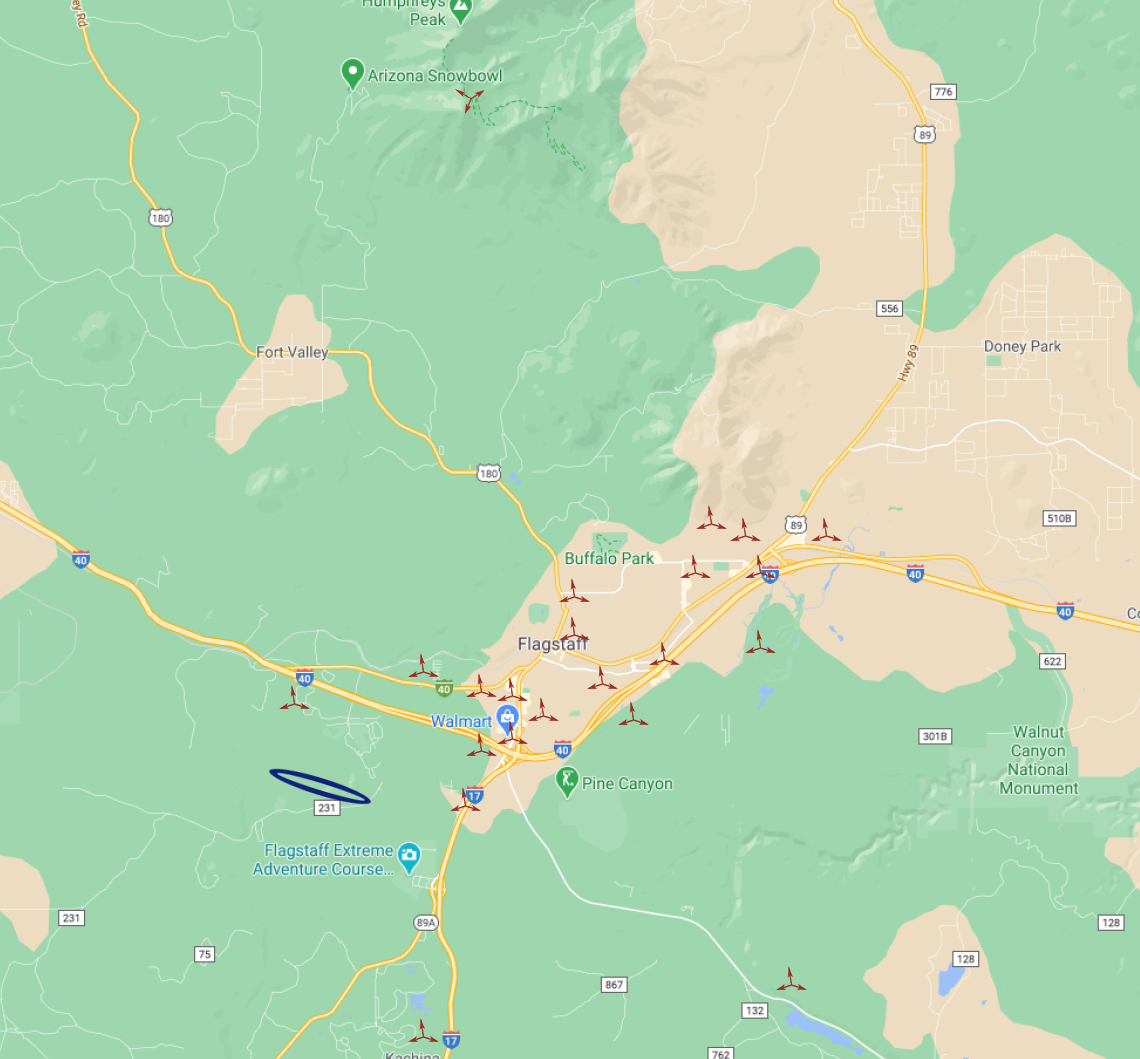}}}
	\subfloat[The captured signals]{\fbox{\includegraphics[width=0.31\linewidth]{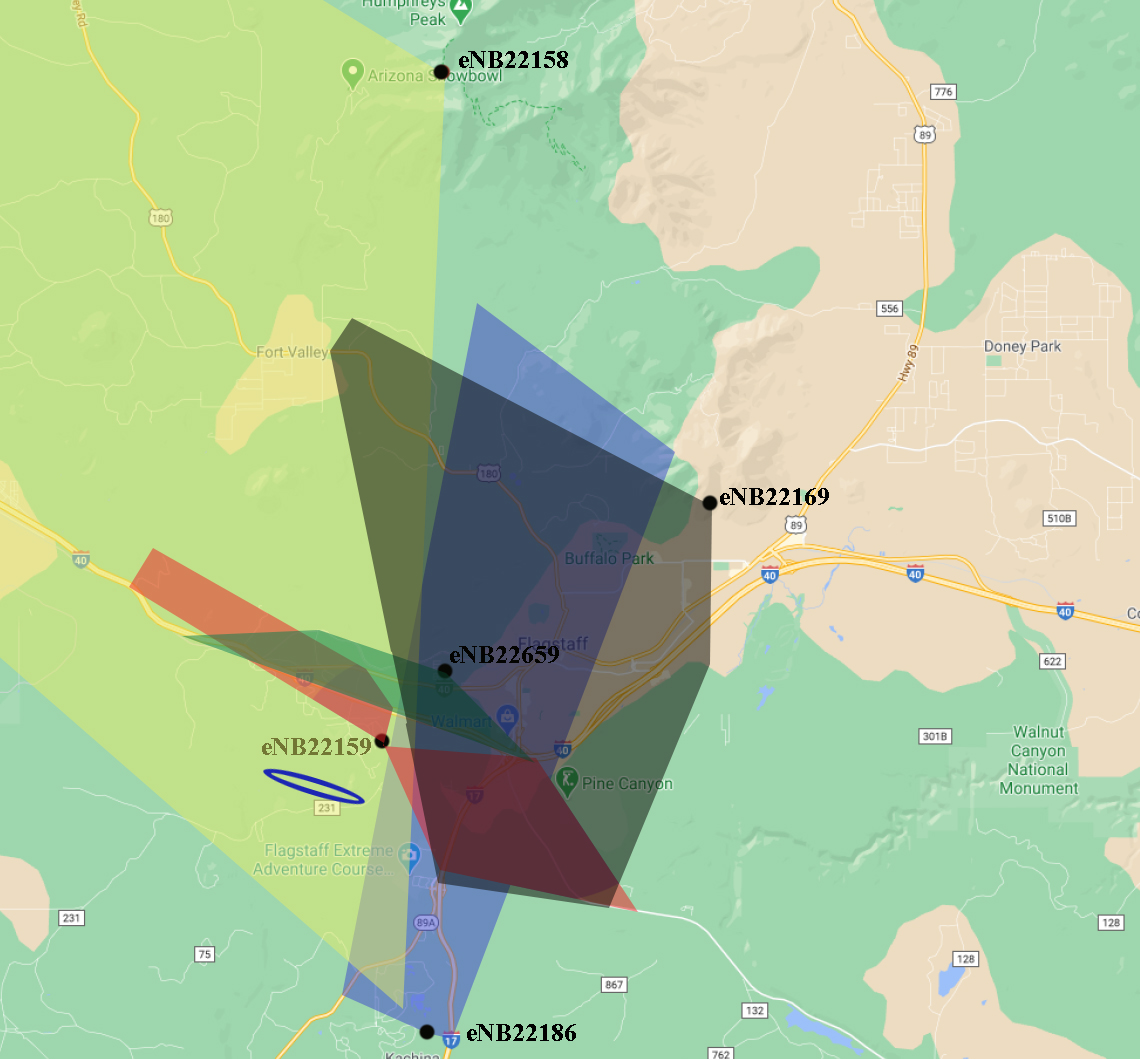}}}
	\subfloat[The test map]{\fbox{\includegraphics[width=0.33\linewidth]{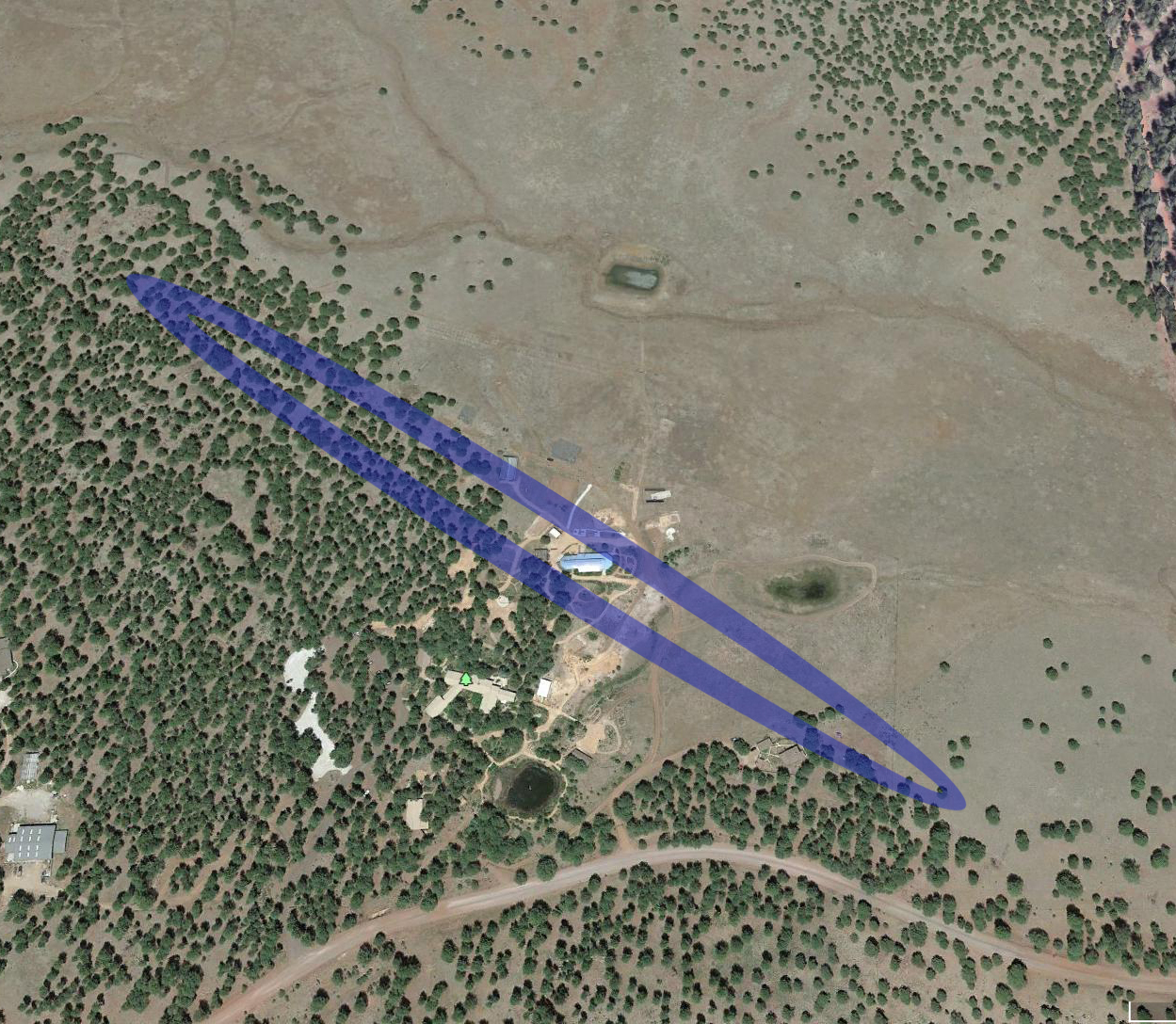}}}
	\caption{A map of the test area.}
	\label{fig::area}
\end{figure*}

In this section, we describe the field flight tests, as well as the data  processing procedure to extract communication parameters from the collected log-files.   We use a commercial drone, DJI Matrice 200, in a rural area near Flagstaff, Arizona, US. Flagstaff lies at approximately 2100 m elevation above sea level. We performed our tests in the Arboretum Garden, one of the southwest experimental garden array (SEGA) sites. The test location is covered by Ponderosa pine trees and surrounded by several Base Transceiver Stations (BTSs). We performed our tests using the commercial Verizon LTE network on its 1700 MHz band. Fig. (\ref{fig::area}) shows the area map. Fig. (\ref{fig::area}a) shows the area with all of the base stations, where Fig. (\ref{fig::area}b) shows only the base stations in which we received their signal in our tests. In this figure, the oval represents the exact test location. Fig. (\ref{fig::area}b) shows also the two-dimension coverage area of the network. This 2D coverage map has been created using multiple drive tests. As shown in this map, the terrestrial user, in the best case, can receive an strong signal from only one base station (the eNB with the identification number 22158), which confirms the weak coverage in this rural area. We put a Samsung S20 phone on the drone with TEMS pocket app \cite{tems} installed on. We flew the drone at three different elevations of 40 m, 80 m, and 120 m, with two different speeds 30 kmph and 60 kmph. While the elevation is limited by the federal aviation administration (FAA) to not exceed 400 feet, i.e. 120 m, the maximum speed is limited by our drone. In each flight instance, the drone take-off from the starting point, flies for 500 m in a straight line and returns back the same way to the take-off point. 

For the measurement process, as we mentioned earlier, we use the TEMS pocket application version 22.1.2 \cite{tems} installed on the attached smart phone. To be able to investigate the communication quality and uplink and downlink data transmission performance, we design a script to call another phone, upload a file to a server, and download a file from another server, all at the same time. We record the RSRP, RSRQ, RSSI, SINR, and uplink and downlink throughput as the performance evaluation metrics. We record the mentioned metrics for the signal of the neighboring cells as well. We also keep the data for all handover processes. We use TEMS discovery to process the files and extract the desired information.       

The collected data, i.e. the output of the TEMS discovery, contains some unnecessary data such as the take-off and landing data when the drone increases or decreases its elevation. Before we use the extracted information, we carefully clean up the data from the unnecessary information. Furthermore, the granularity of TEMS discovery in managing the data is 2 seconds. We represent the data based on the location of drone, starting from the starting point, going in the straight line for 500 m, in steps of 50 m. To calculate the information at the exact locations of interest, we interpolated the required data points, using Lagrange interpolation.      

%% file: DataAnalysis.tex
\section{Experimental Results}
\label{sec::dataAnalysis}

\begin{figure}[h!]
	\centering
	\subfloat[Different elev.  (speed=60kmph)]{\includegraphics[width=.5\linewidth]{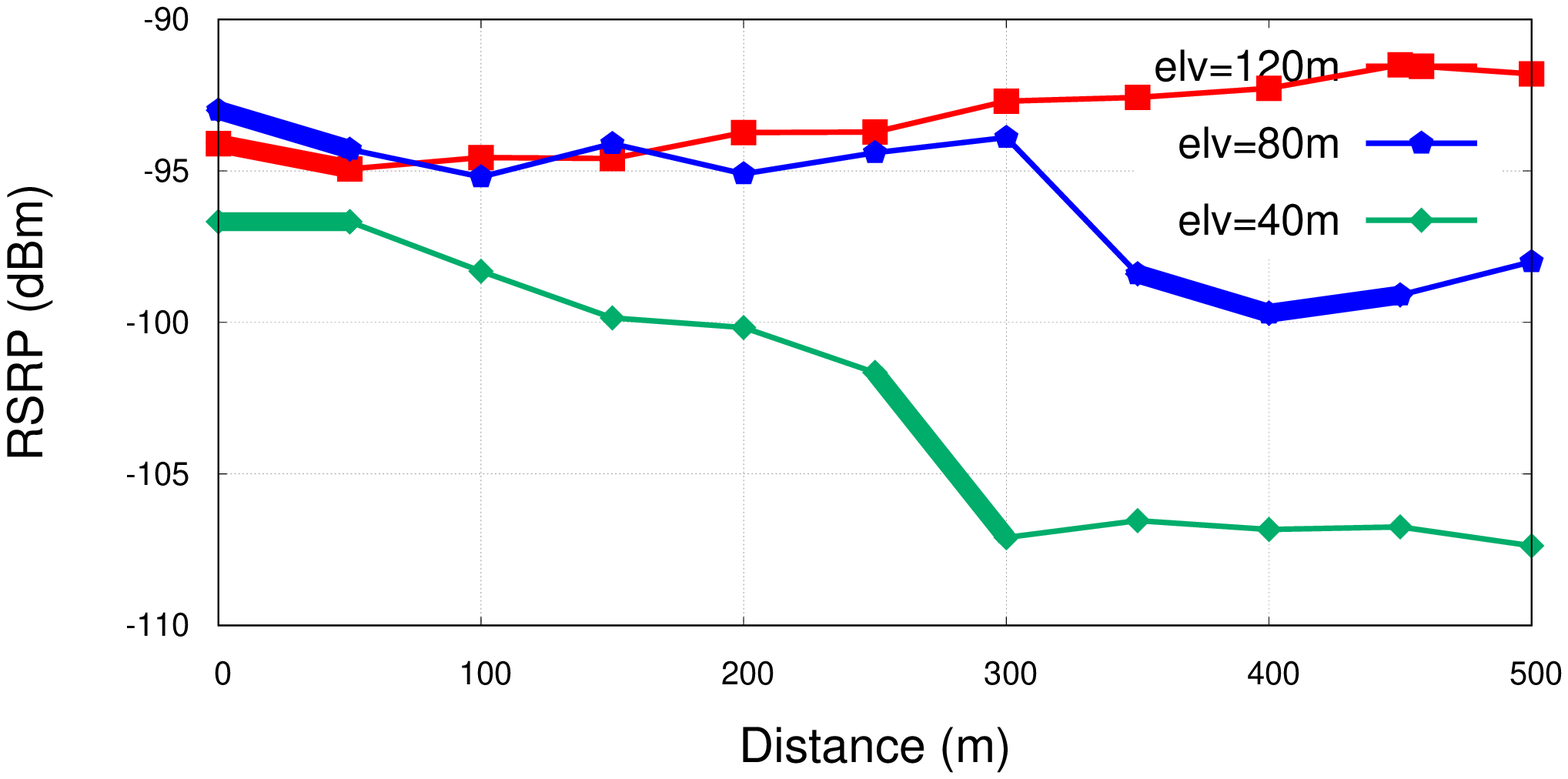}}
	\subfloat[Different elev. opposite direction (speed=60kmph)]{\includegraphics[width=.5\linewidth]{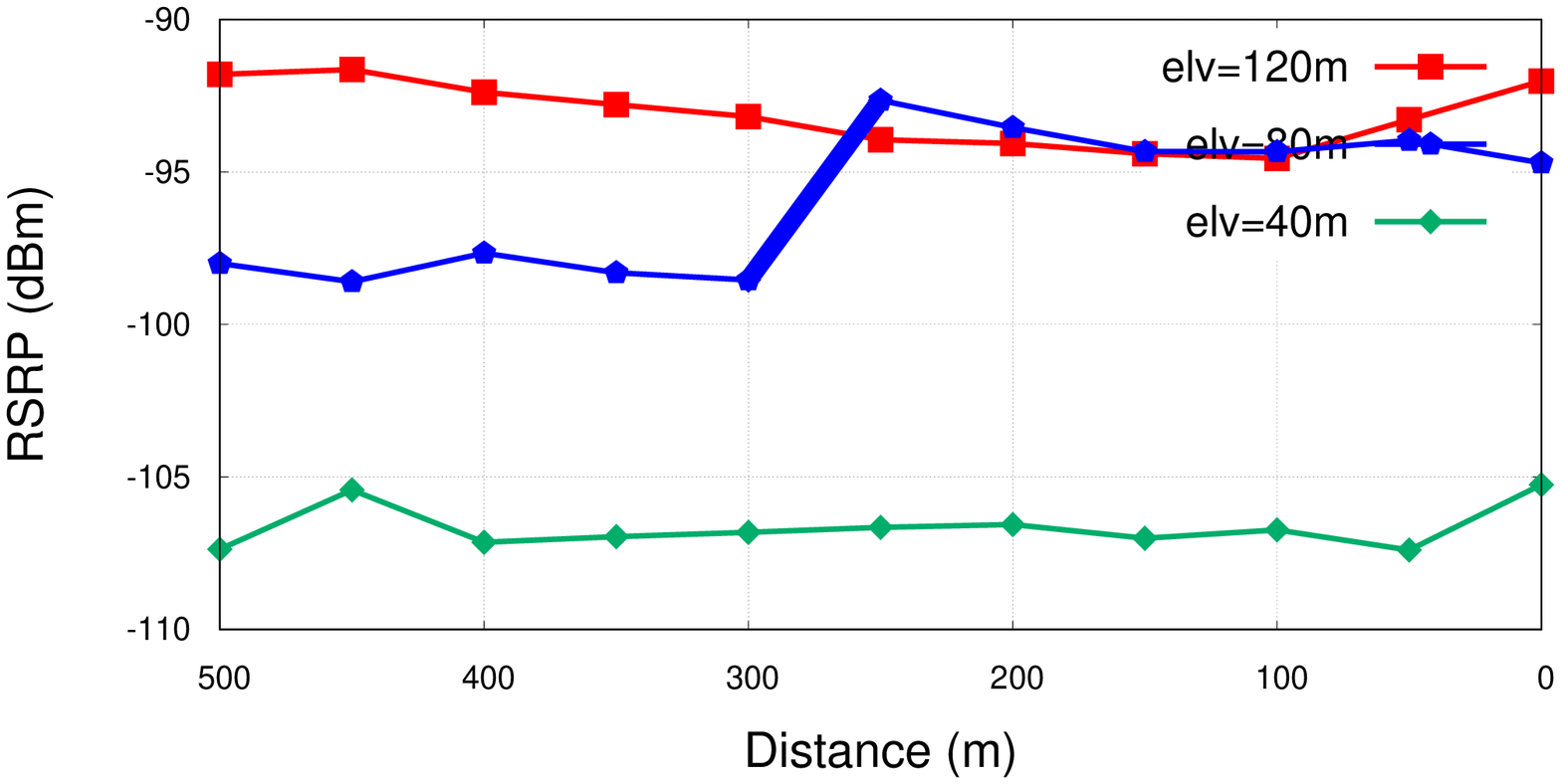}}\\
	\subfloat[Different speeds (elev=120m)]{\includegraphics[width=.5\linewidth]{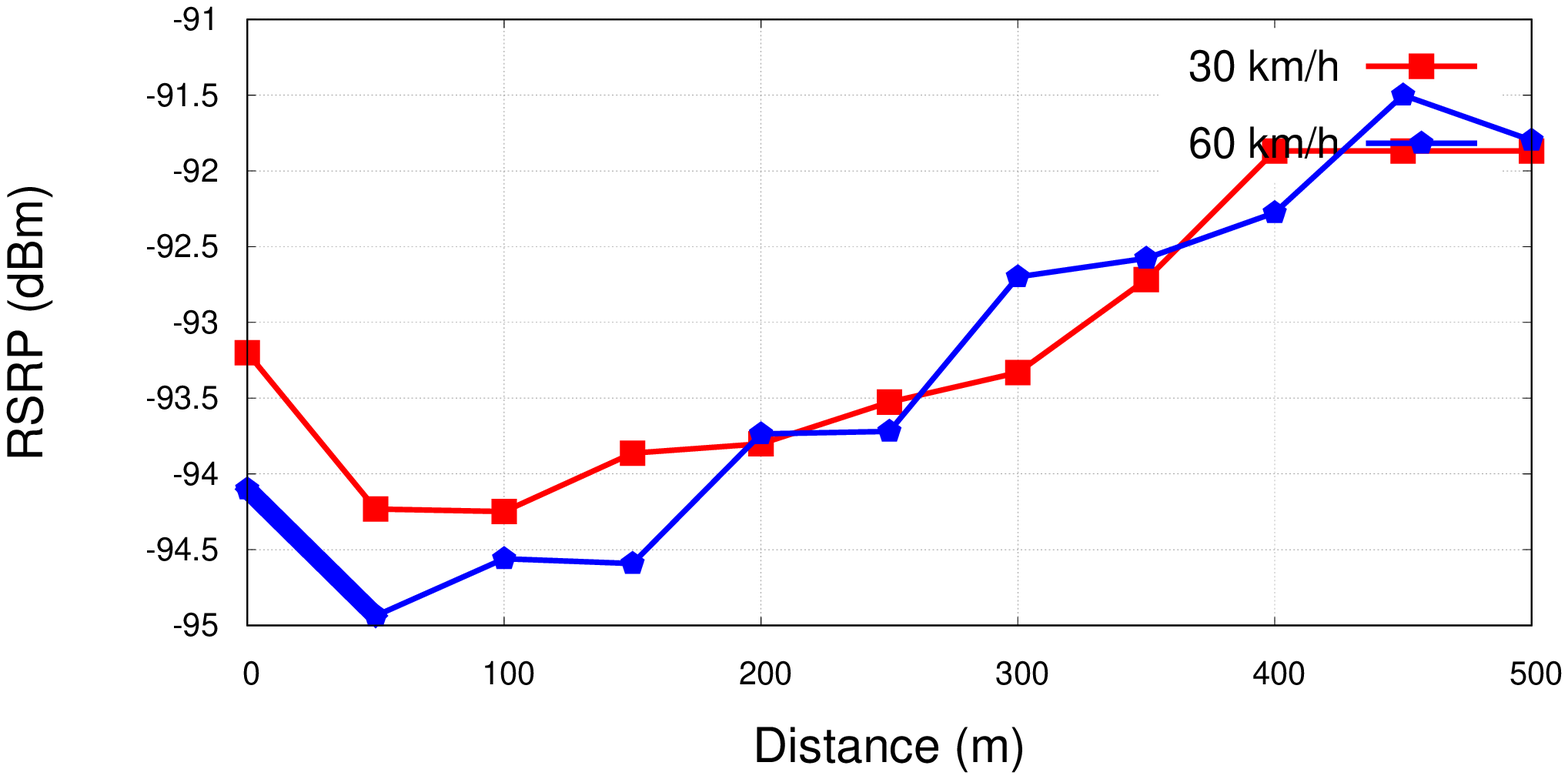}}
	\subfloat[Different speeds opposite direction(elev=120m)]{\includegraphics[width=.5\linewidth]{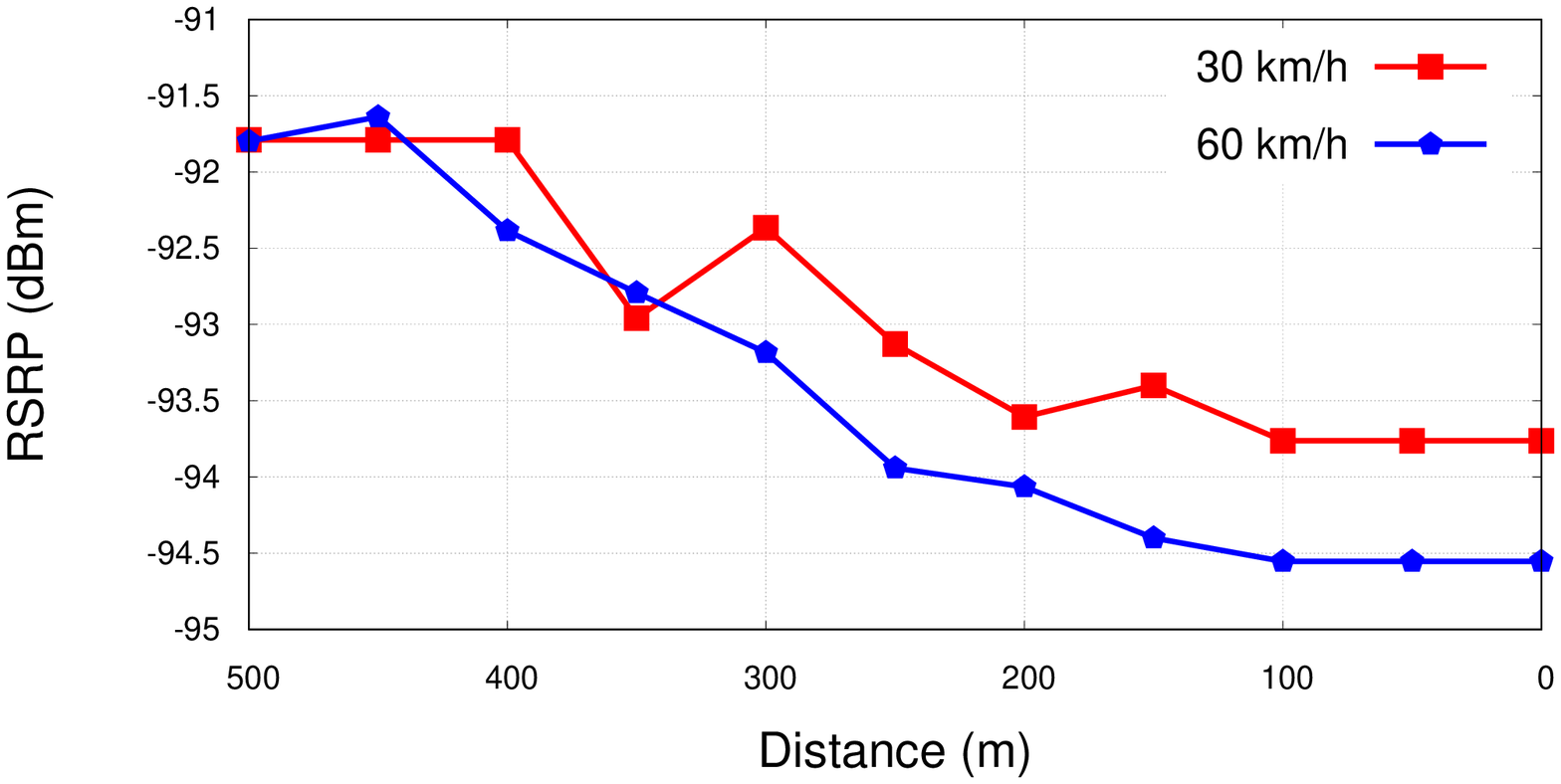}}\\
	\caption{A comparison of RSRP for different elevations and different UAV speeds.}
	\label{fig::RSRP}
\end{figure}

In this section, we first represent and analyze the measured metrics regarding the serving cell signal. The measured and analyzed metrics are RSRP, RSRQ, RSSI, SINR, downlink throughput, and uplink throughput. We then compare the RSRP and RSRQ between the serving cell signal and the signals of the neighboring cells.  
The Reference signal received power (RSRP) is the first investigated metric for the serving cell signal, in our measurement and analysis. According to 3GPP, RSRP is defined as the "linear average over the power contributions of the resource elements that carry cell-specific reference signals within the considered measurement frequency bandwidth" \cite{3GPP}. In a simpler form, the RSRP represents the power of the reference signal at the receiver in a LTE network, excluding the noise and interference from neighboring cells. It is measured in dBm, and the signal is considered excellent if $(RSRP\ge -80 dBm)$, good if $(-90 dBm\le RSRP\le -80 dBm)$, fair to poor if $(-100 dBm\le RSRP\le -90 dBm)$, and no signal if $(RSRP  \le -100 dBm)$. Fig. (\ref{fig::RSRP}) shows the RSRP for the combination of different speeds and different elevations and for both ways of going from the starting point to 500 m away, and returning back to the starting point. The highlighted intervals represent a handover process happened at that interval. 

Figures (\ref{fig::RSRP}a and b) show the RSRP for the fixed speed of 60 kmph and three different elevations of 40 m, 80 m, and 120 m, for both directions. As it is clear from this figure, the higher elevation helps the network node to get a signal with higher power, as the higher elevation leads to more areas with a dominant line-of-sight signal. The number of handover processes shows also obvious decrements at the elevation of 120 m. Fig. (\ref{fig::RSRP}c,d) show the same results for the elevation of 120m and different speeds of 30 kmph and 60 kmph. While the two speeds represent results with an almost identical pattern, the lower speed shows slightly better performance. The RSRP shows almost the same pattern of results for the back and forward directions. Thus, for the other metrics, we represent the figures only for one way.

\begin{figure}[h!]
	\centering
	\subfloat[Different elev. (speed=60 kmph)]{\includegraphics[width=.5\linewidth]{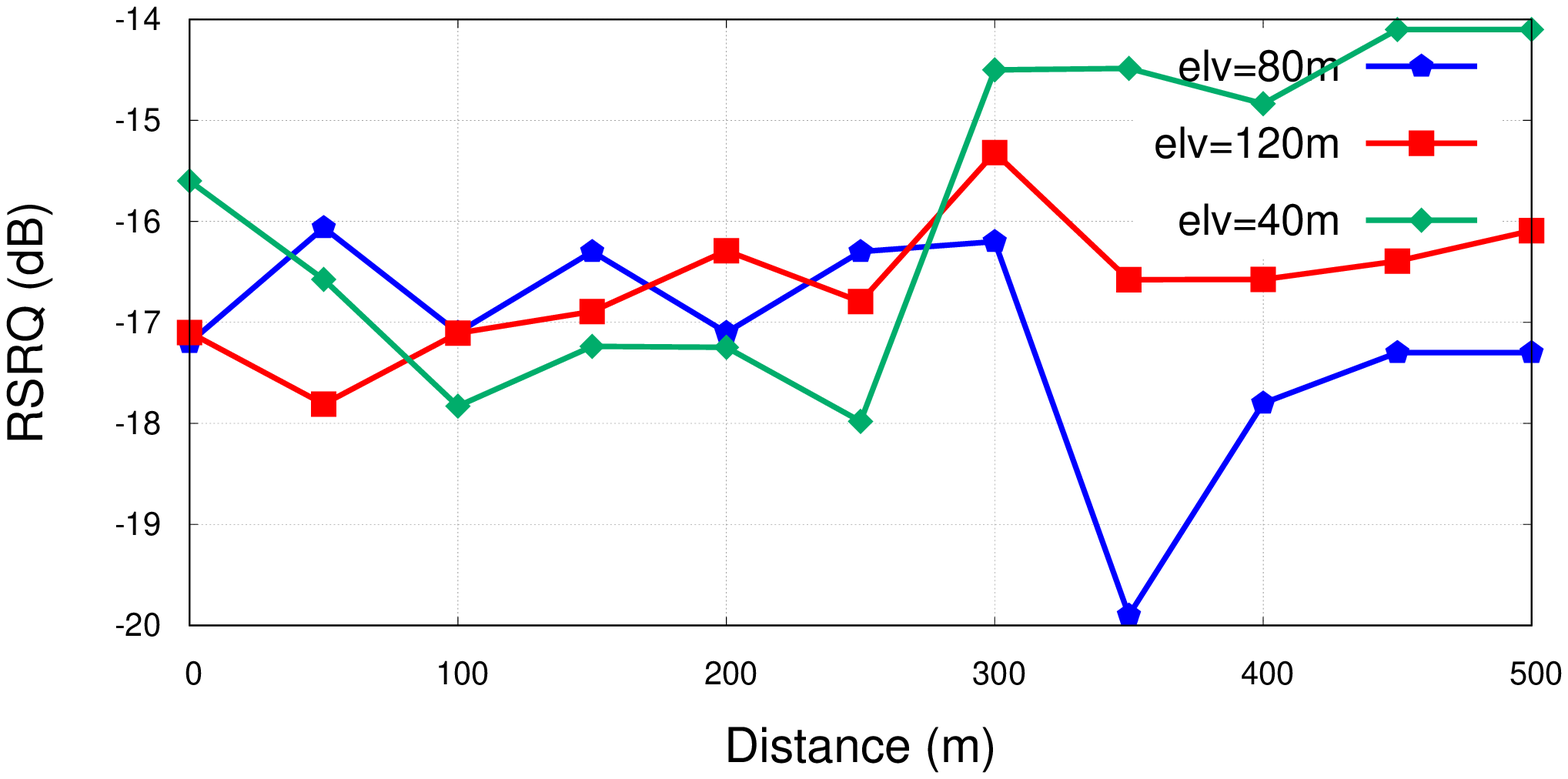}}
	\subfloat[Different speeds (elev=120 m)]{\includegraphics[width=.5\linewidth]{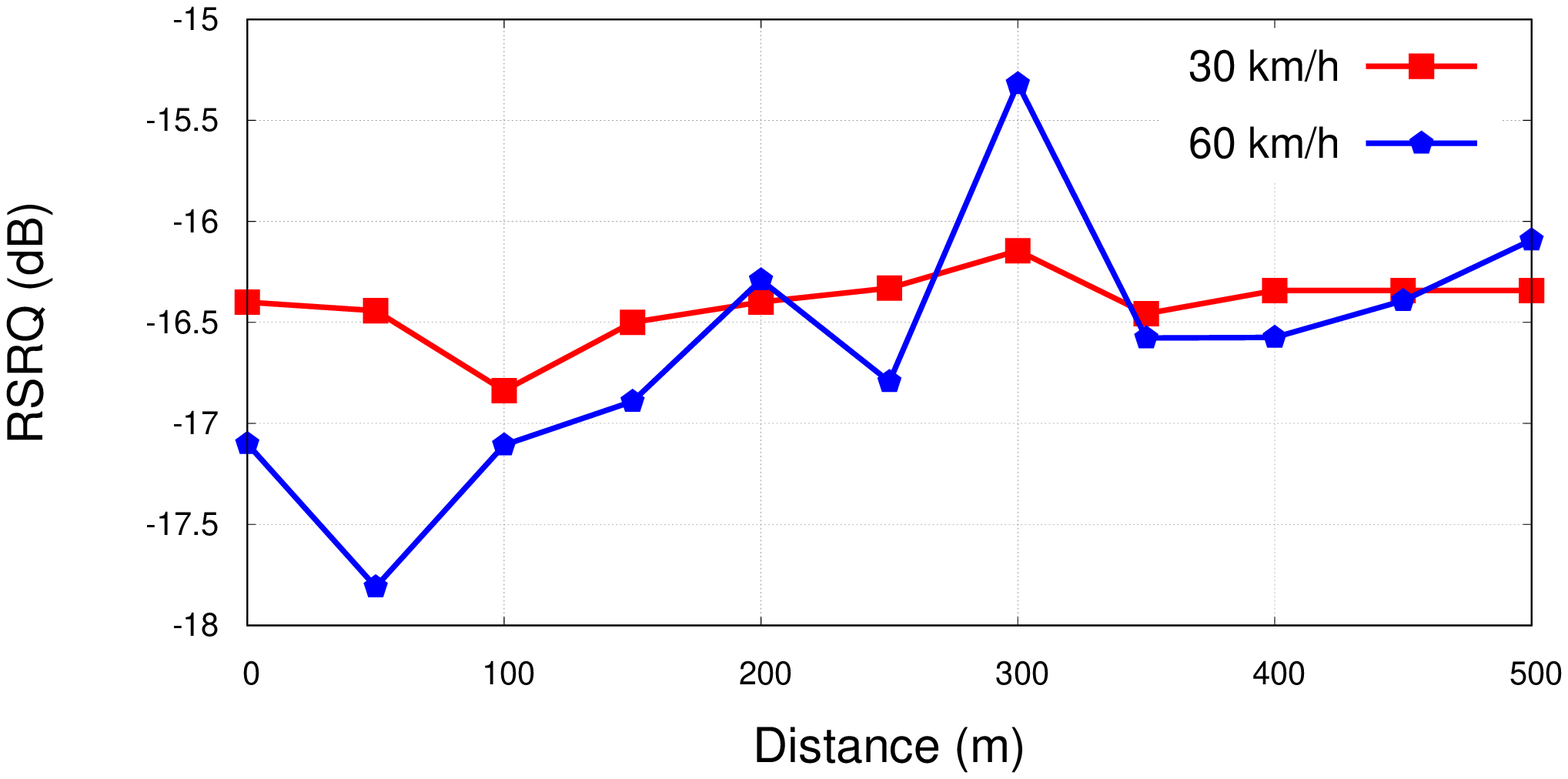}}
	\caption{A comparison of RSRQ for different elevations and different UAV speeds.}
	\label{fig::RSRQ}
\end{figure}

Reference signal received quality (RSRQ) is the next investigated metric. The RSRQ represents the quality of received signal at the user equipment and it is measured in dB. While the RSRP is the main metric for decision-making on handover and cell reselection, it can provide additional information when RSRP is insufficient. The signal is considered  as excellent if $(RSRQ\ge -10 dB)$, good if $(-15dB\le RSRQ \le -10dB)$, fair to poor if $(-20dB\le RSRQ \le -15dB)$, and no signal if ($RSRQ\le -20dB$). Fig. (\ref{fig::RSRQ}) shows the serving signal RSRQ for the combination of different elevations and different speeds. Looking at the range of variation of this metric, we find that there is not a significant improvement among different settings. However, the higher elevation and lower speed show slightly better performance in terms of RSRQ.  

\begin{figure}[h!]
	\centering
	\subfloat[Different elev. (speed=60 kmph)]{\includegraphics[width=.5\linewidth]{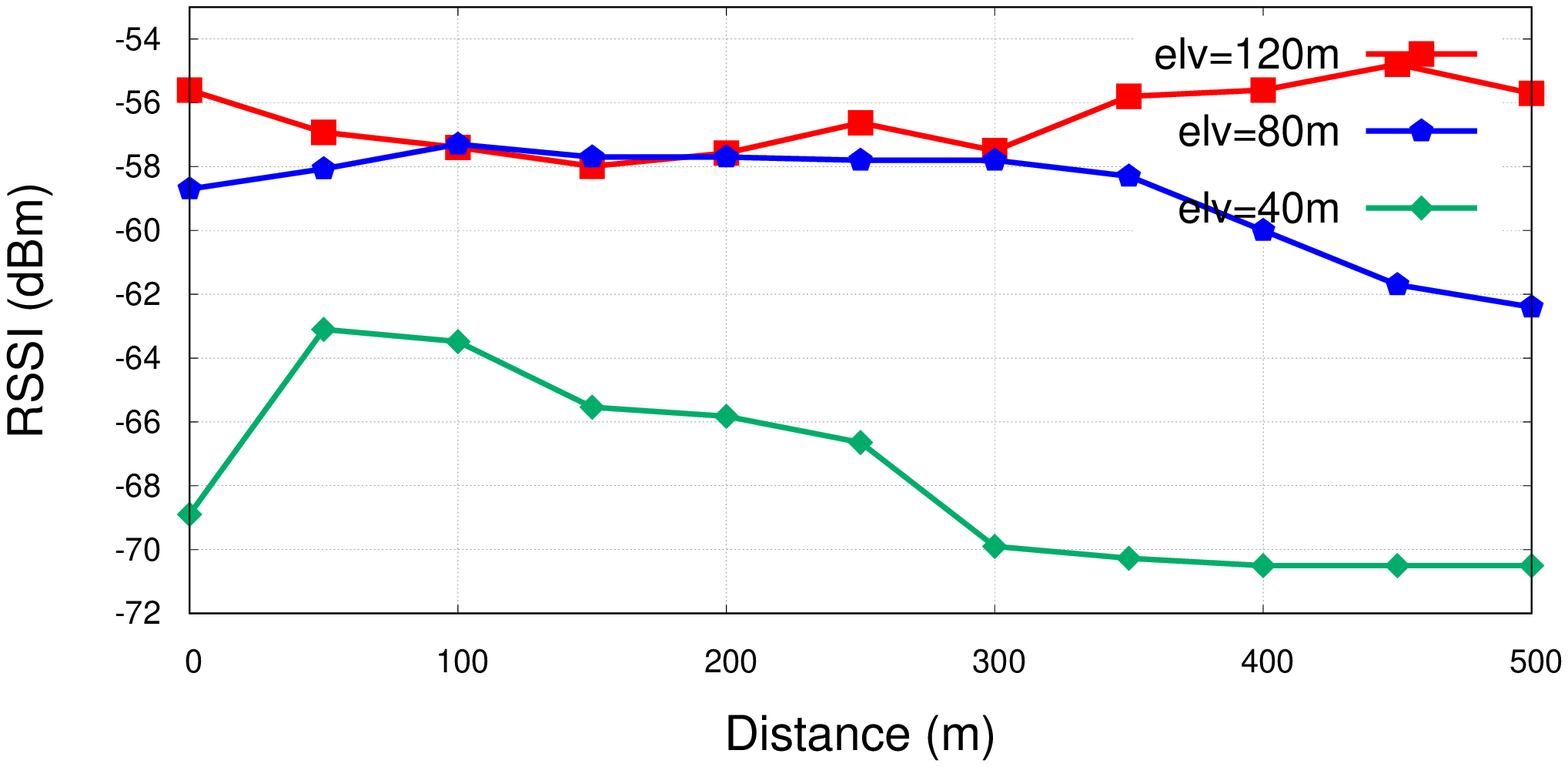}}
	\subfloat[Different speeds (elev=120 m)]{\includegraphics[width=.5\linewidth]{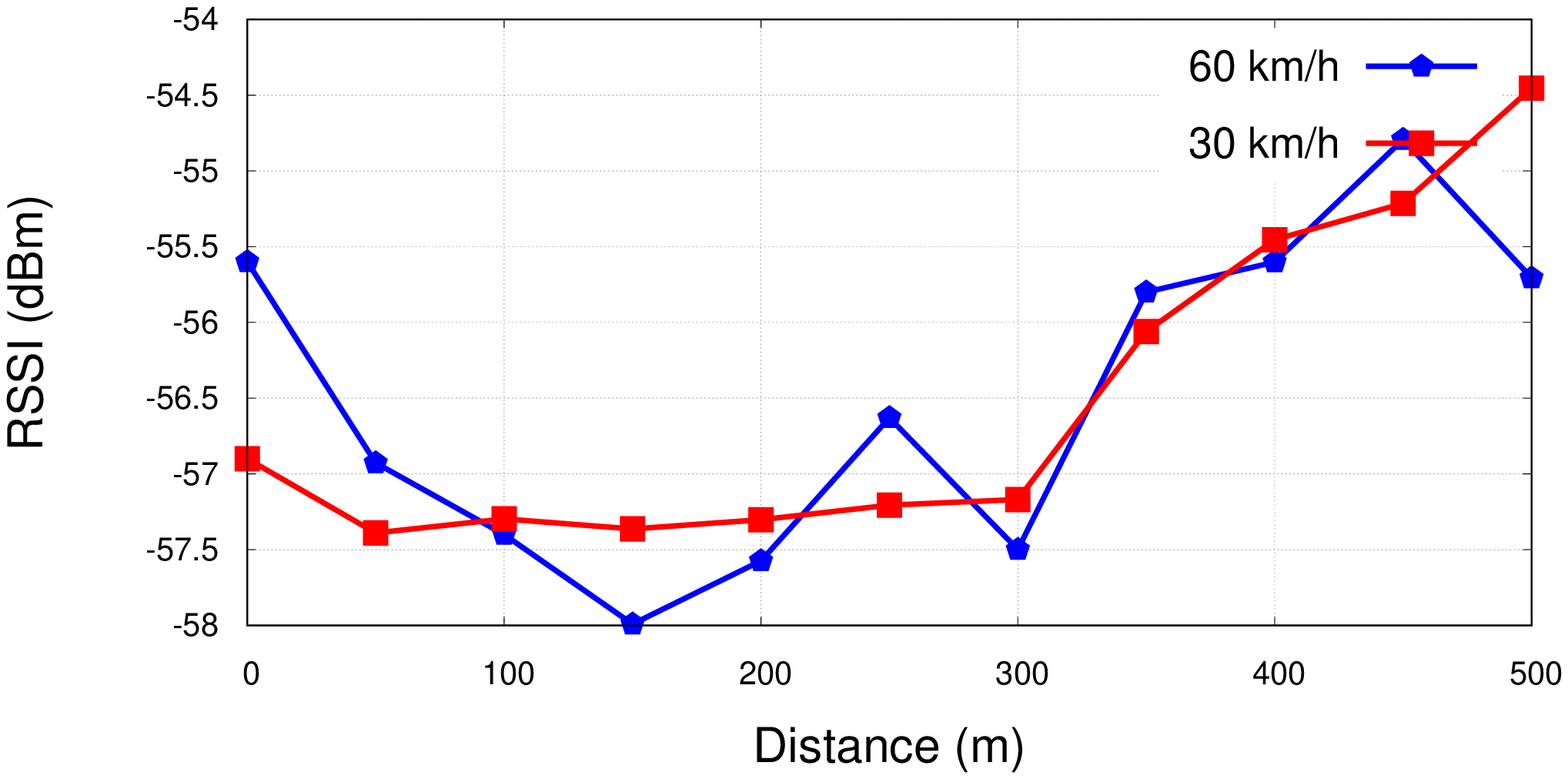}}
	\caption{A comparison of RSSI for different elevations and different UAV speeds.}
	\label{fig::RSSI}
\end{figure}

The Received signal strength indicator (RSSI) is the next measured metric. The RSSI represents the strength of the received signal, considering the noise and interference, and it is measured in dBm. Fig. (\ref{fig::RSSI}) shows the RSSI for different elevations and different speeds. Again the superiority of the higher elevations is significant for this metric. As we discussed earlier, the test is done in a rural area covered by trees. That is why when the flight altitude is much larger than the trees' height, the signal strength remains almost the same in the low-altitude tests. However, we note that for the same elevation, the speed has no obvious effect on this parameter.   

\begin{figure}[h!]
	\centering
	\subfloat[Different elev. (speed=60 kmph)]{\includegraphics[width=.5\linewidth]{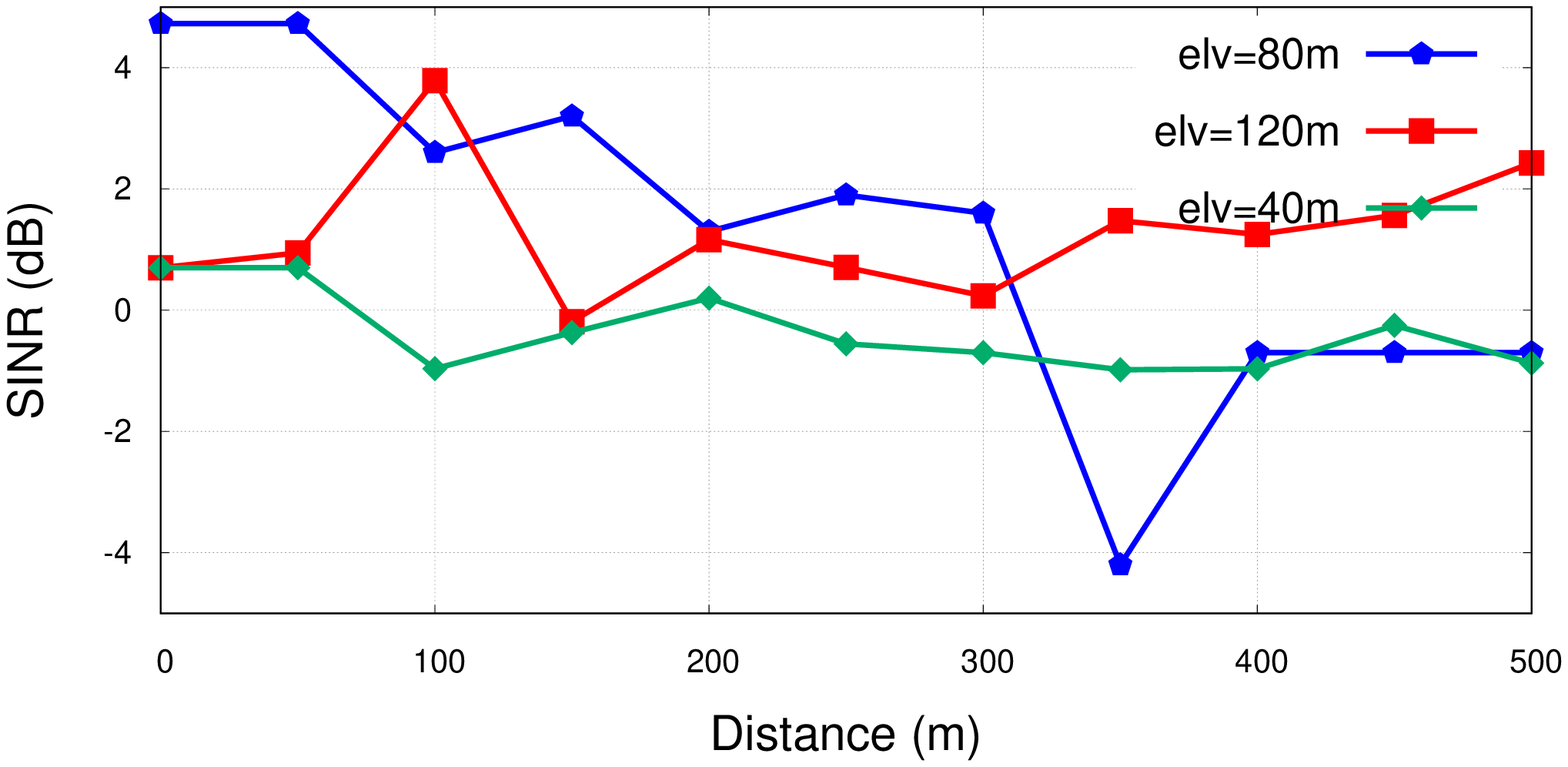}}
	\subfloat[Different speeds (elev=120 m)]{\includegraphics[width=.5\linewidth]{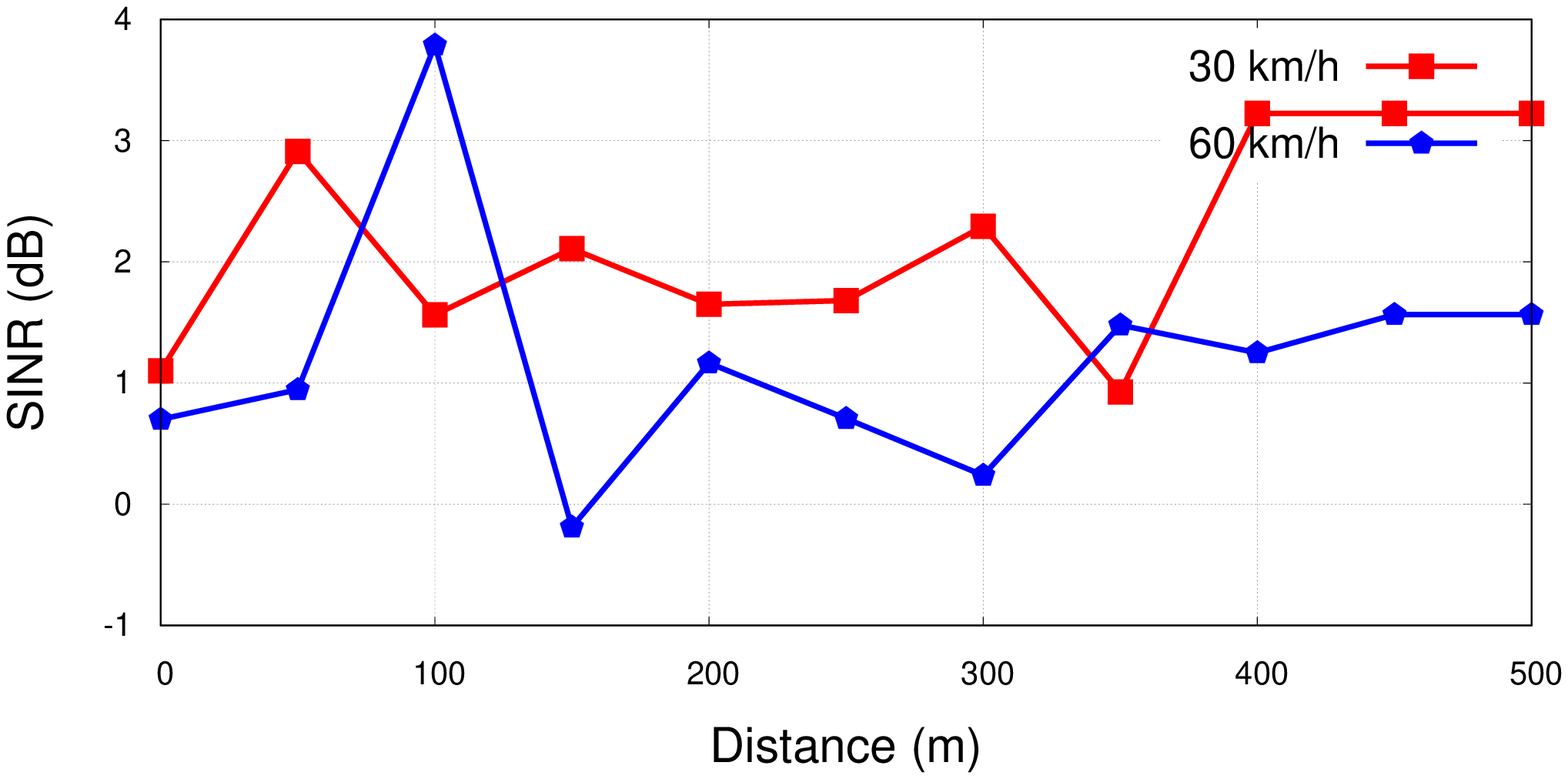}}
	\caption{A comparison of SINR for different elevations and different UAV speeds.}
	\label{fig::SINR}
\end{figure}

Signal to interference and noise ratio (SINR), measured in dB, is the next measured metric. This metric is defined as the RSRP divided by the sum of the interference power from the neighboring cells and noise power. SINR is important as a quantifying metric of the relationship between the radio frequency and achievable throughput. Fig. (\ref{fig::SINR}) shows the SINR results for different elevations and different UAV speeds. While the higher speed as well as the lowest elevation, in most cases, lead to lower SINR, for the elevations of 80 m and 120 m there is no significant superiority in terms of SINR.  

The last investigated metrics are the achievable Transmission Control Protocol (TCP) downlink and uplink throughput. To measure the throughput, in the script which is running on the smartphone, we start uploading a large file to a server with a bandwidth much higher than that of cellular network. At the same time, we download a large enough file, again from a server with a bandwidth much more than that of cellular network. In this case, the limiting parameter is the maximum achievable throughput at the user equipment side, which is the network throughput. 

\begin{figure}[t!]
	\centering
	\subfloat[Different elev. (speed=60 kmph)]{\includegraphics[width=.5\linewidth]{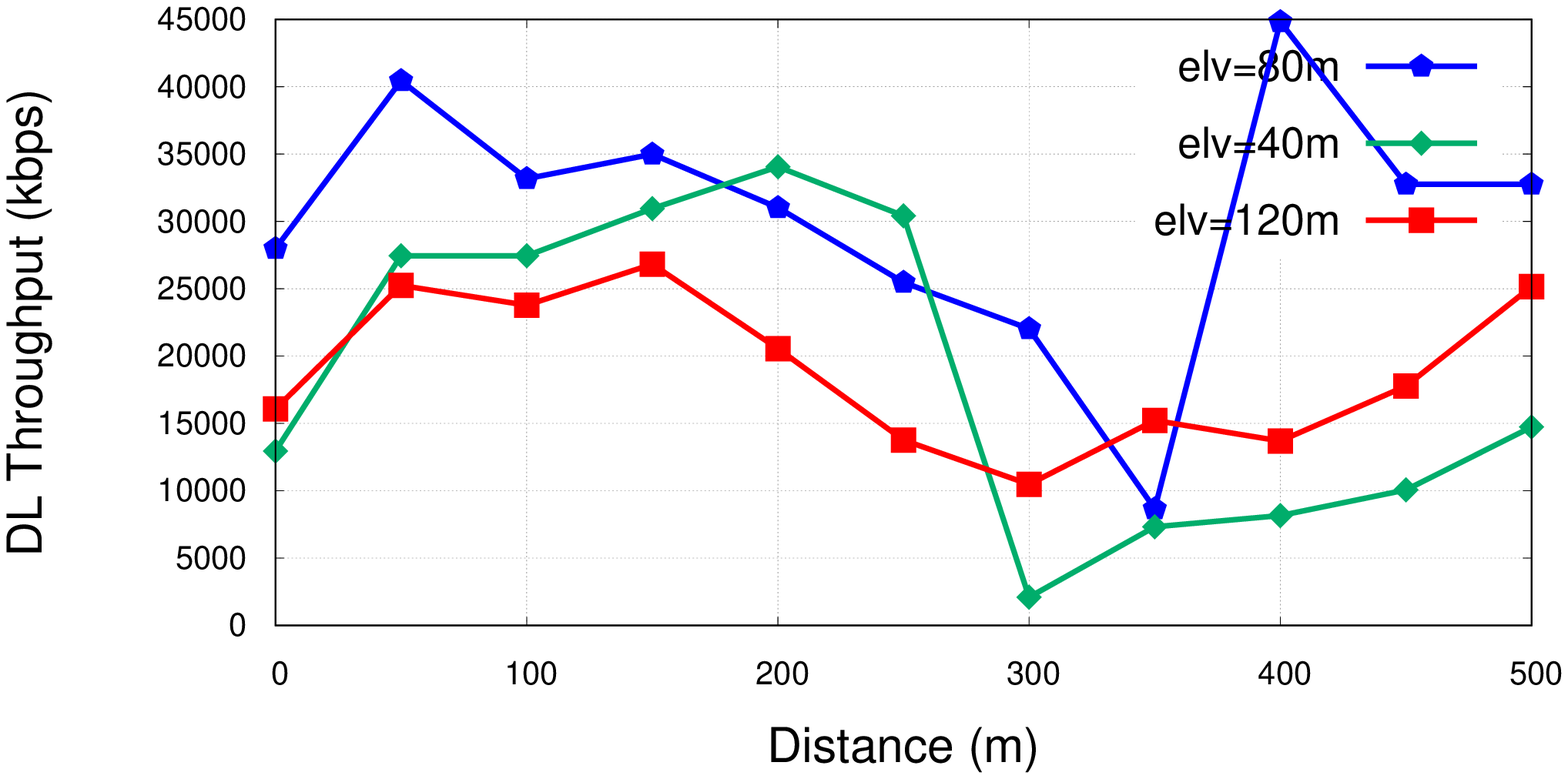}}
	\subfloat[Different speeds (elev=120 m)]{\includegraphics[width=.5\linewidth]{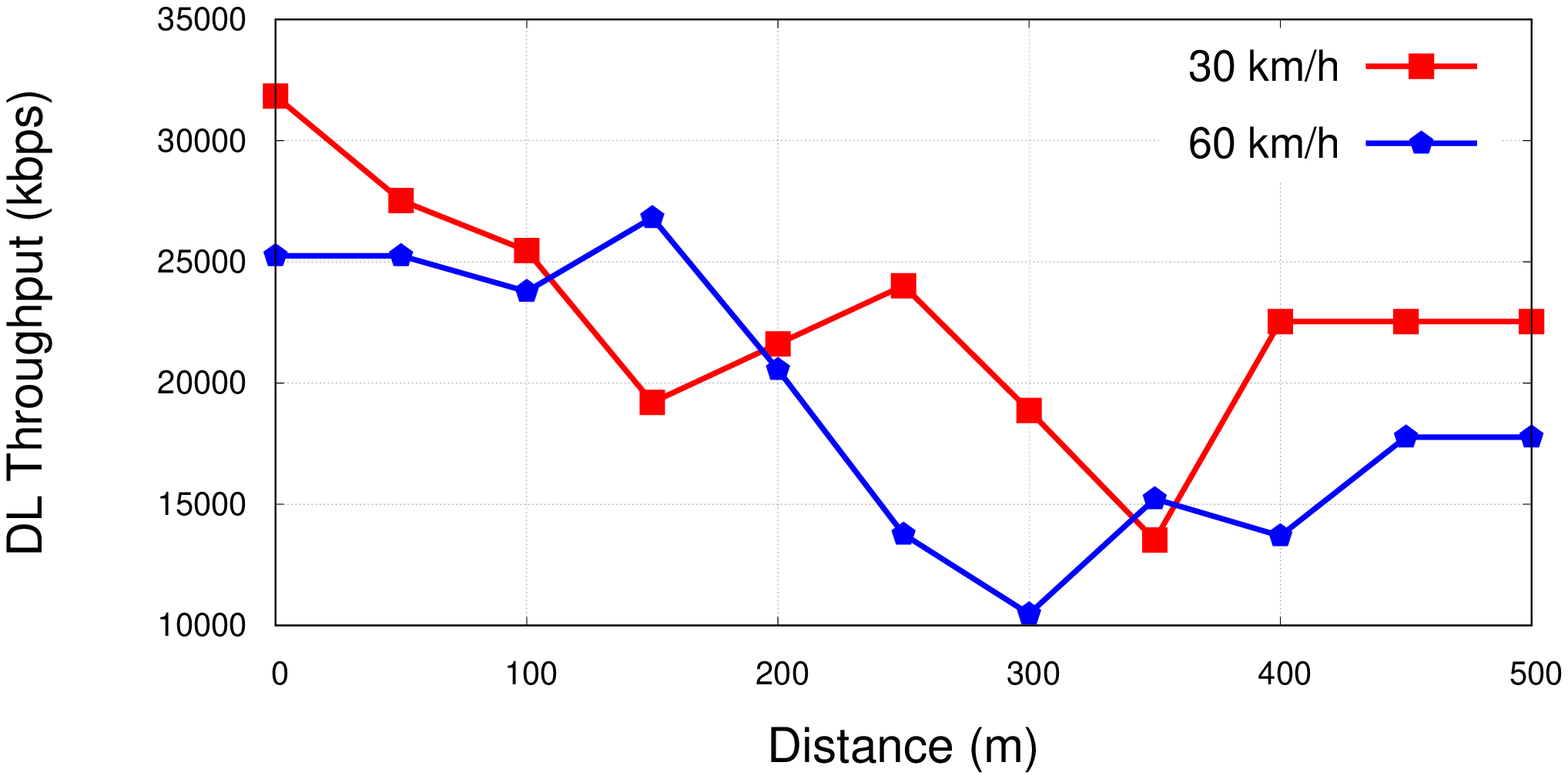}}
	\caption{A comparison of downlink throughput for different elevations and different UAV speeds.}
	\label{fig::dlThroughput}
\end{figure}

\begin{figure}[t!]
	\centering
	\subfloat[Different elev. (speed=60 kmph)]{\includegraphics[width=.5\linewidth]{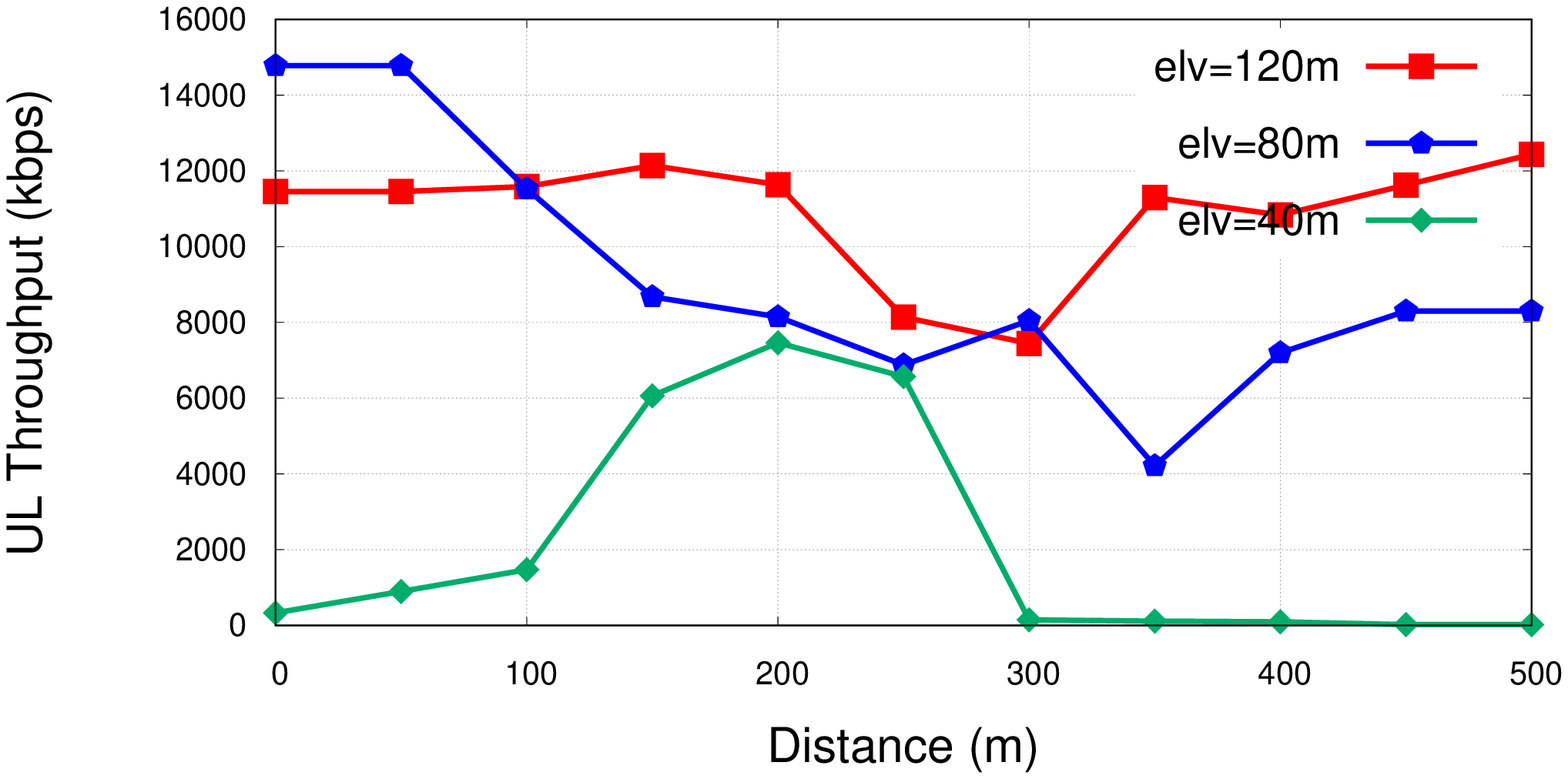}}
	\subfloat[Different speeds (elev=120 m)]{\includegraphics[width=.5\linewidth]{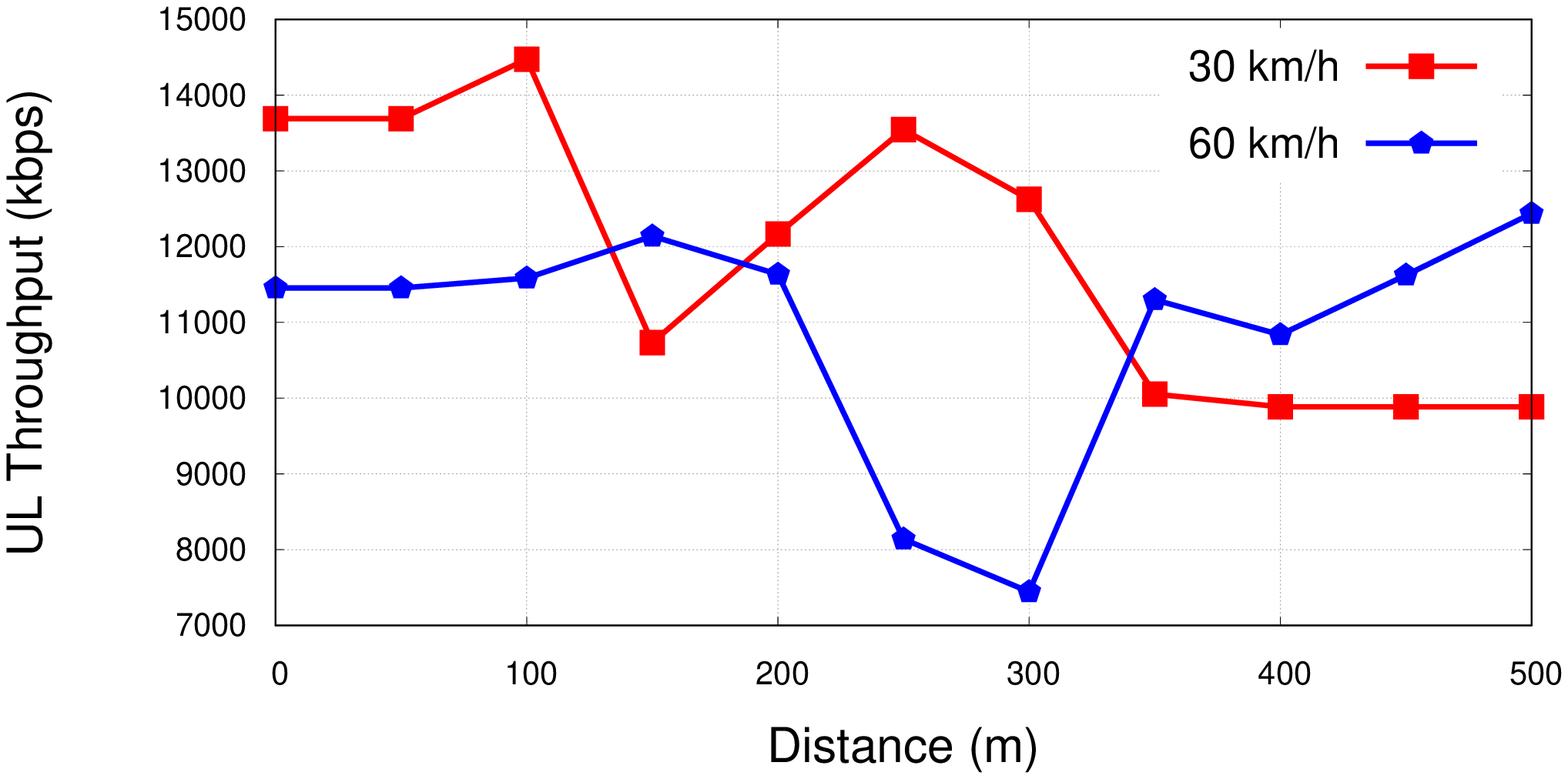}}
	\caption{A comparison of uplink throughput for different elevations and different UAV speeds.}
	\label{fig::ulThroughput}
\end{figure}

Fig. (\ref{fig::dlThroughput} and \ref{fig::ulThroughput}) show the throughput for a combination of different elevations and different drone speeds for downlink and uplink, respectively. In Fig. (\ref{fig::dlThroughput}), the variation in the elevation does not show a significant change in the achievable downlink throughput as the lower speed shows its superiority. However, the elevation of 80 m shows slightly better downlink throughput, in most points. One possible reason could be the higher interference at the higher elevation and the absence of line of sight signals at  the lower elevation. In Fig. (\ref{fig::ulThroughput}), we see that the uplink shows an obvious superiority at the higher elevation and the lower speed flight.

\begin{figure*}[t]
	\centering
	\subfloat[Elevation= 40 m ]{\includegraphics[width=.33\linewidth]{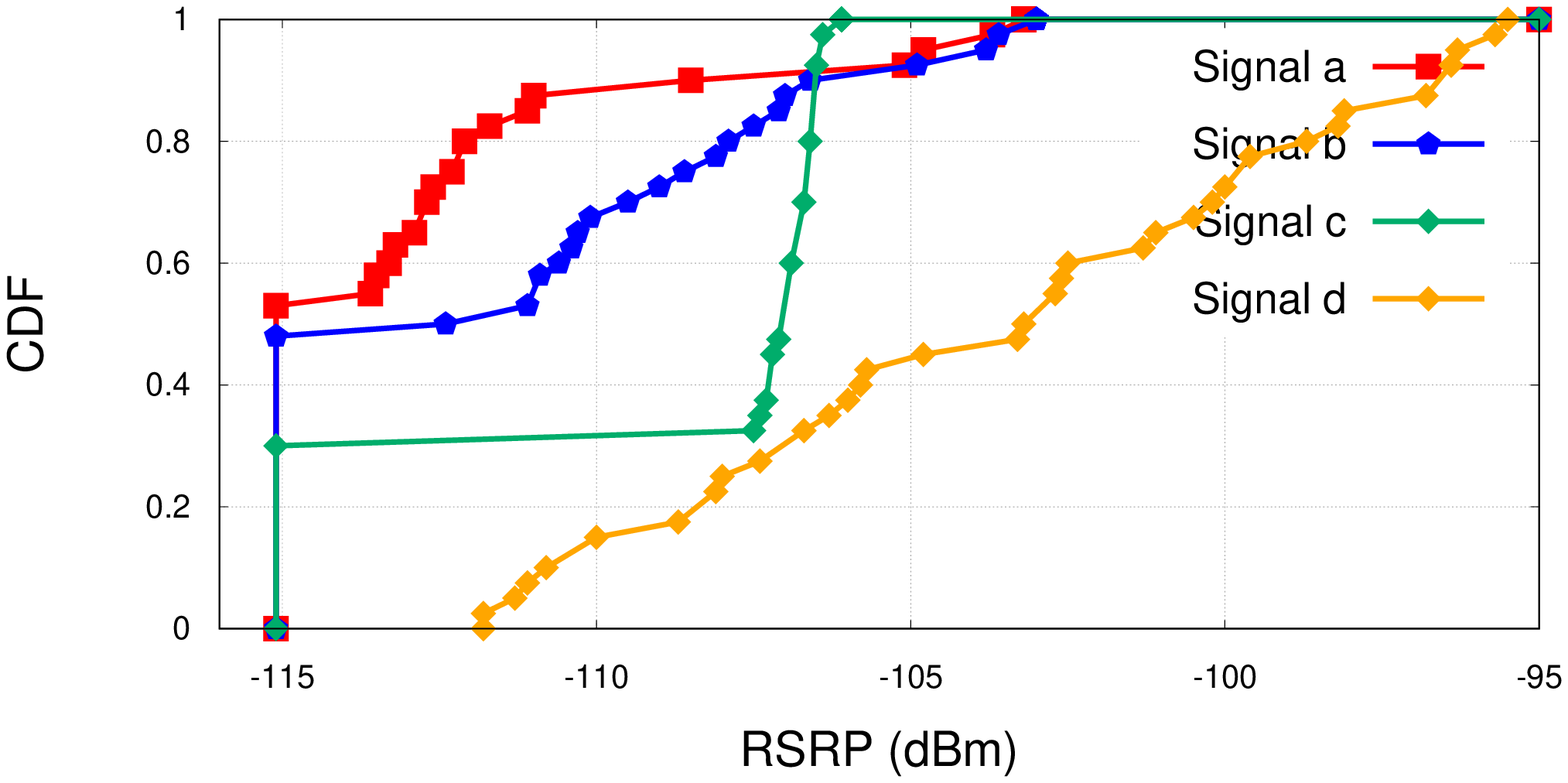}}
	\subfloat[Elevation= 80m]{\includegraphics[width=.33\linewidth]{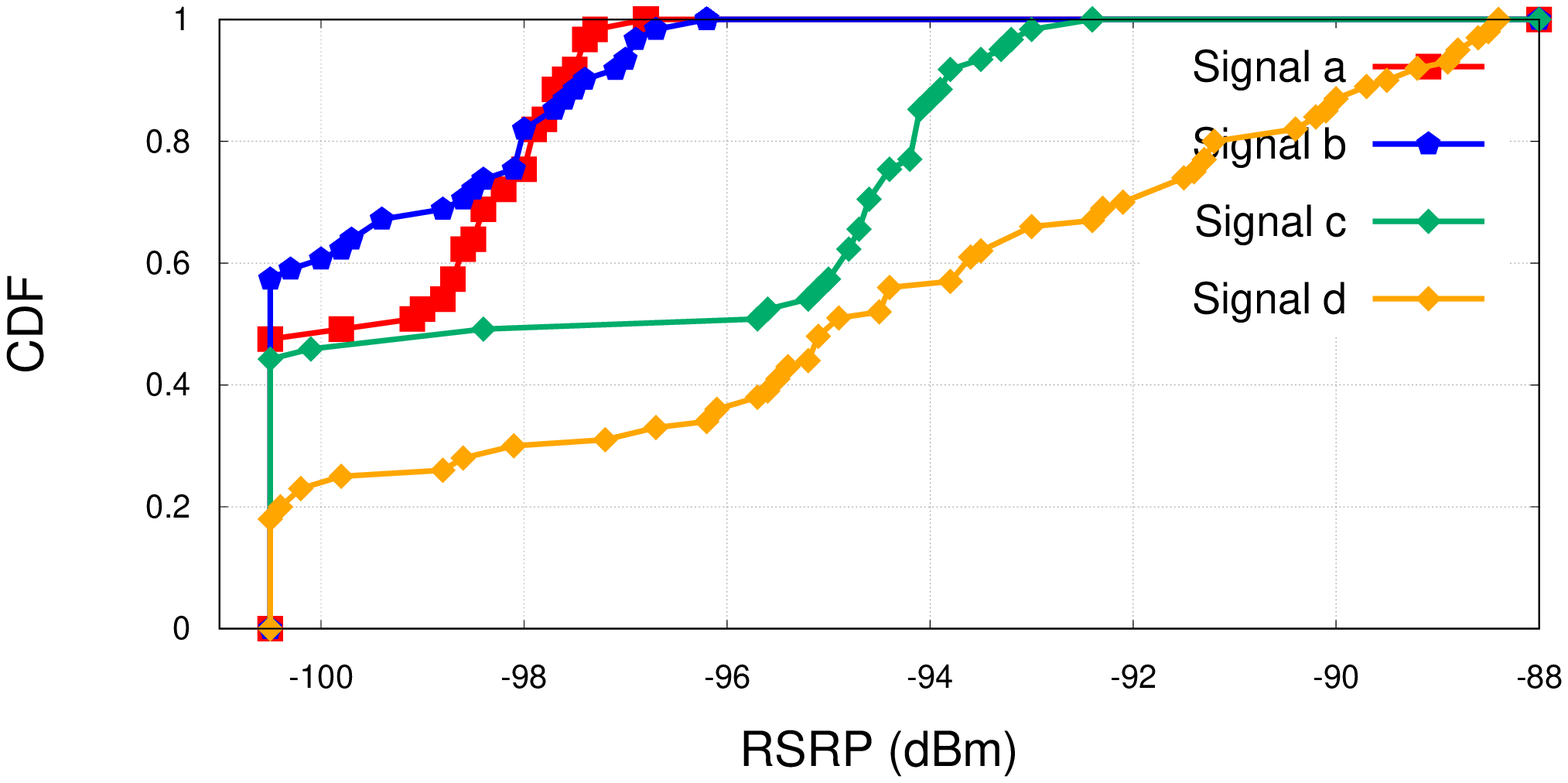}}
	\subfloat[Elevation= 120 m]{\includegraphics[width=.33\linewidth]{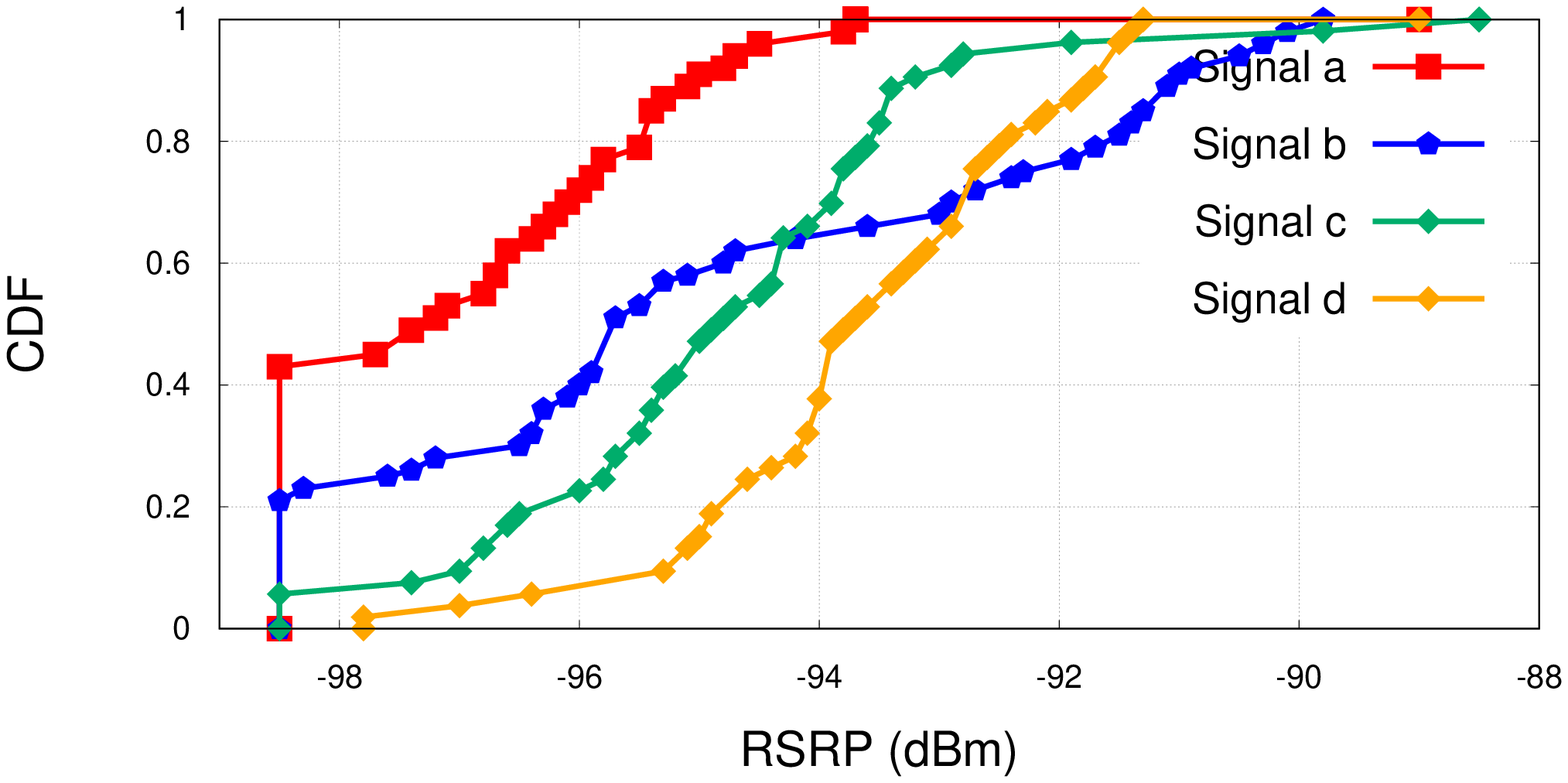}}
	\caption{A comparison of RSRP CDF for signals with the highest RSRP in different elevations and a fixed speed of 60 kmph. }
	\label{fig::RSRPcdfElv}
\end{figure*}

Now, we compare the RSRP and RSRQ of the best available signals including the serving cell's signal and the signal of the three of its neighboring cells. Fig. (\ref{fig::RSRPcdfElv}) shows the empirical cumulative distribution function (CDF) for the RSRP of the signals with the highest RSRP for three different elevations and a fixed speed of 60 kmph. In Fig. (\ref{fig::RSRPcdfElv}a), the signal 'a' always has the close RSRP to the other signals 'b' and 'c' but was never been choosen as the serving cell. In Fig. (\ref{fig::RSRPcdfElv}b), signals 'b', 'c', and 'd' serve as the main signal, where signal 'a' always was from a neighboring cell. In Fig. (\ref{fig::RSRPcdfElv}b), signal 'd', in almost all times, was the serving cell signal. 

\begin{figure*}[t]
	\centering
	\subfloat[Speed= 30 kmph ]{\includegraphics[width=.33\linewidth]{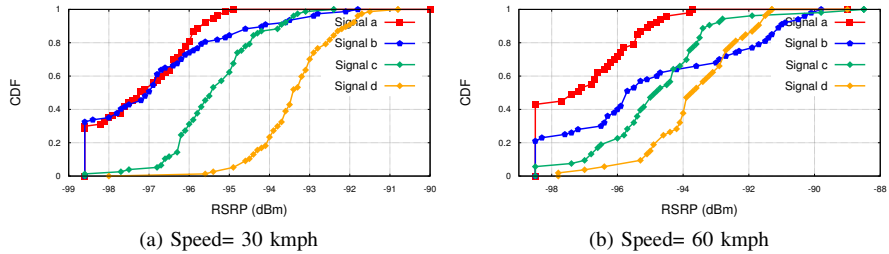}}
	\subfloat[Speed= 60 kmph]{\includegraphics[width=.33\linewidth]{figs/RSRPcdf120elv60}}
	\caption{A comparison of RSRP CDF for signals with the highest RSRP in different speeds and a fixed elevation of 120m.}
	\label{fig::RSRPcdfSpeed}
\end{figure*}

\begin{figure*}[t]
	\centering
	\subfloat[Elevation= 40 m ]{\includegraphics[width=.33\linewidth]{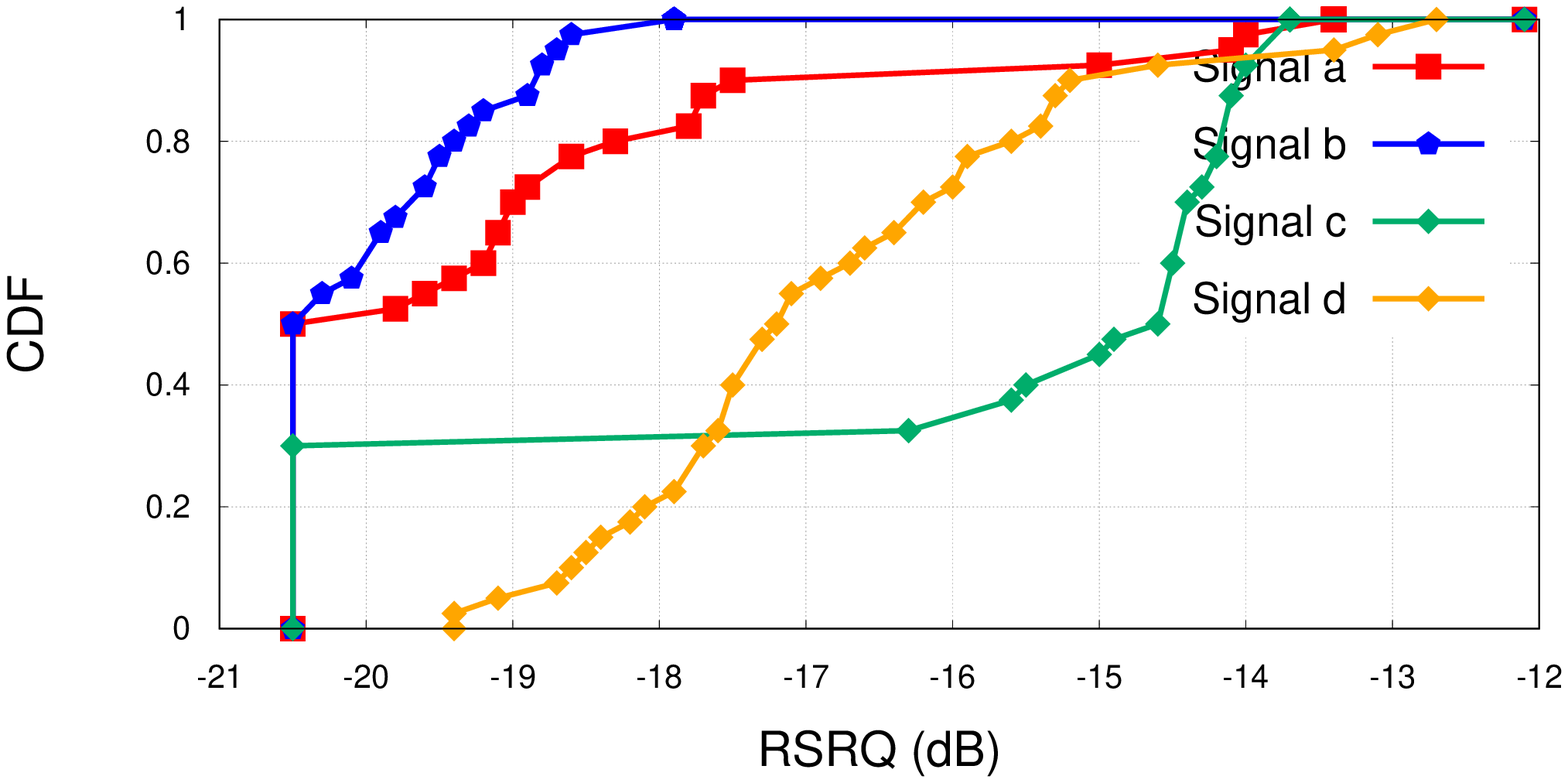}}
	\subfloat[Elevation= 80 m]{\includegraphics[width=.33\linewidth]{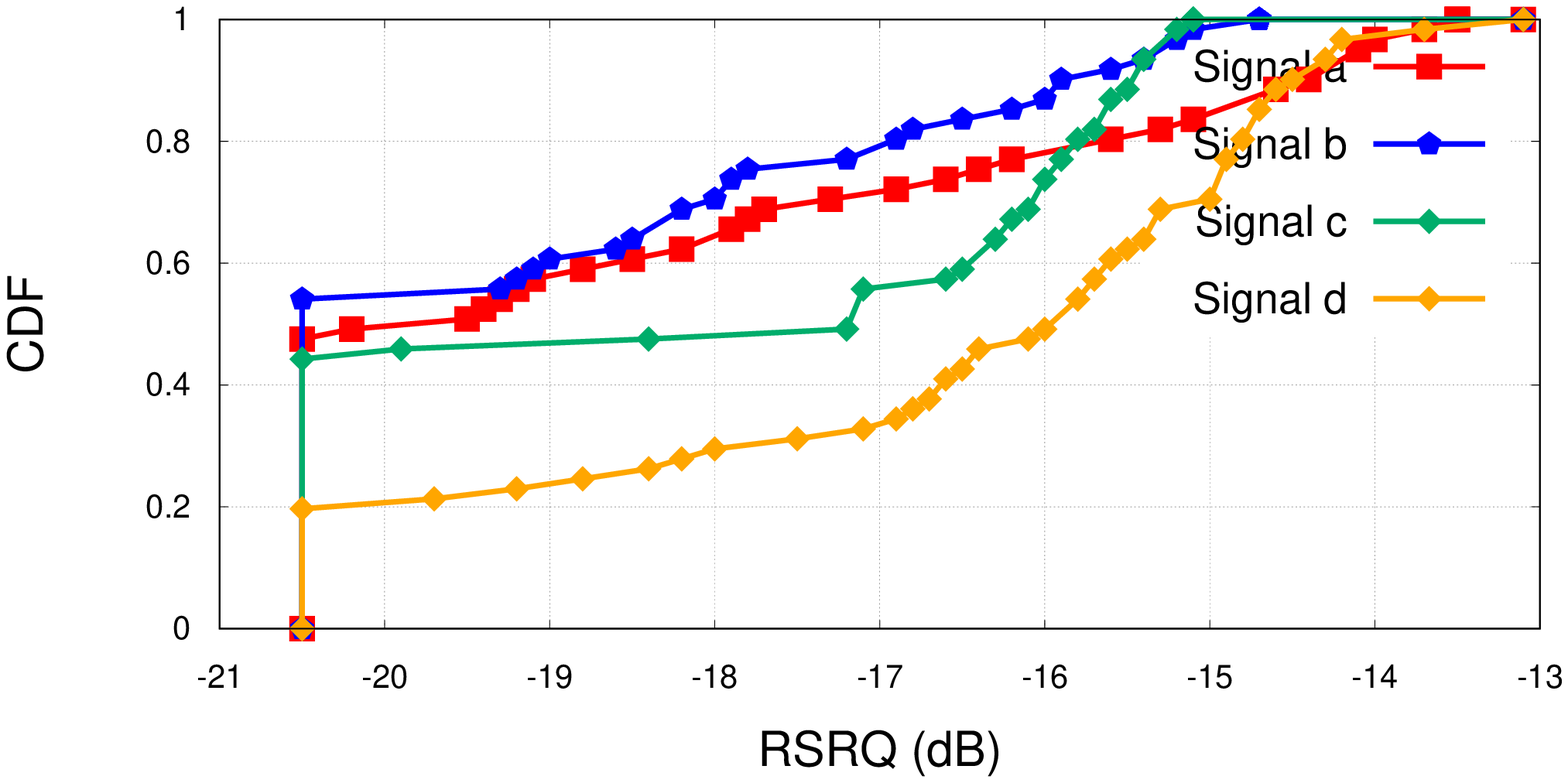}}
	\subfloat[Elevation= 120 m]{\includegraphics[width=.33\linewidth]{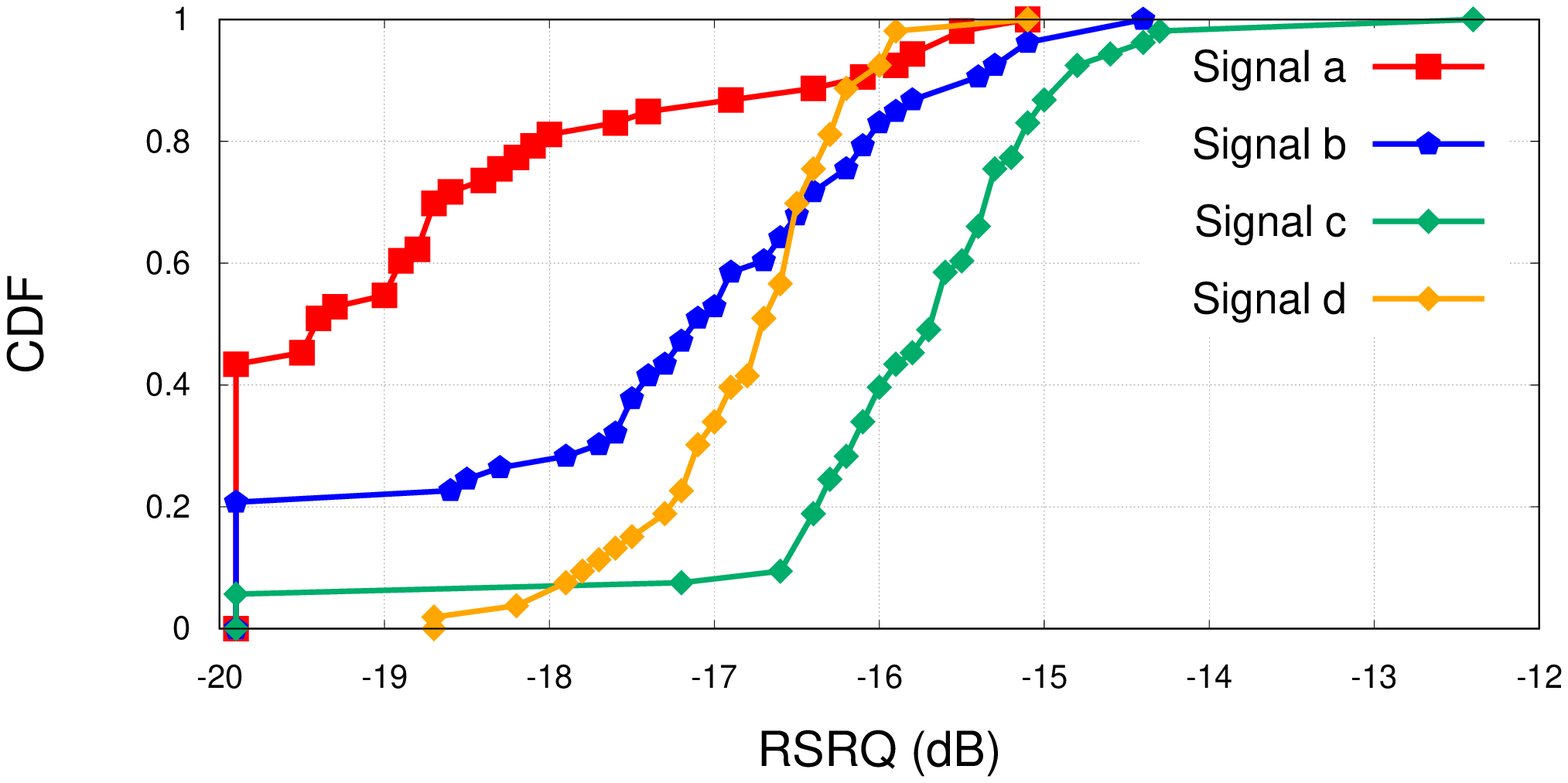}}
	\caption{A comparison of RSRQ CDF for signals with the highest RSRQ in different elevations and a fixed speed of 60 kmph.}
	\label{fig::RSRQcdfElv}
\end{figure*}

Fig. (\ref{fig::RSRPcdfSpeed}) shows the empirical CDF for the RSRP of the signals with the highest RSRP for two different speeds and an elevation of 120 m. In both figures of (\ref{fig::RSRPcdfSpeed}- a and \ref{fig::RSRPcdfSpeed}- b) signal 'd' is chosen as the serving cell signal. While in Fig. (\ref{fig::RSRPcdfSpeed}a), it is always the signal with the highest RSRP, in Fig. (\ref{fig::RSRPcdfSpeed}b), signals 'b' and 'c' somewhere act better but never have been chosen as the serving cell signals. It is worthy to mention that we did not experience any call drop during the tests. As it is clear from Fig. (\ref{fig::RSRPcdfSpeed} and \ref{fig::RSRQcdfElv}), most of the times, there are at least a couple of signals with enough power to keep the call alive. However, we see a different number of handovers in different test settings. Table (\ref{tbl::HO}) compares the number of handover processes for different test settings. This table shows that at the elevation of 120 m, we experience less number of handovers in comparison with other elevations, where the elevation of 80 m leads to the highest number of handover processes. Generally, we find that the highest speed leads to a slightly less number of handover processes. It seems that at the elevation of 40 m, there are a couple of handover processes due to the environmental obstacles. At the elevation of 80 m, the interference of other signals seems to have its highest effect. However, at the elevation of 120 m, the effect of interference from other cells' signals has its lowest effect on the serving signal, which leads to a less often handovers.

\begin{table}[t!]
\caption{Number of handovers}
\begin{center}
\begin{tabular}{|c|c|c||c|c|}
    \hline
    \multicolumn{5}{|c|}{Number of Handovers}\\\hline
    \multicolumn{3}{|c||}{Elevation (m)}&\multicolumn{2}{|c|}{Speed (kmph)}\\\hline
    40&80&120&30&60\\\hline
    2&4&1&1&0\\
    \hline
\end{tabular}
\end{center}
\label{tbl::HO}
\end{table}

\begin{figure*}[t]
	\centering
	\subfloat[Speed= 30 kmph ]{\includegraphics[width=.33\linewidth]{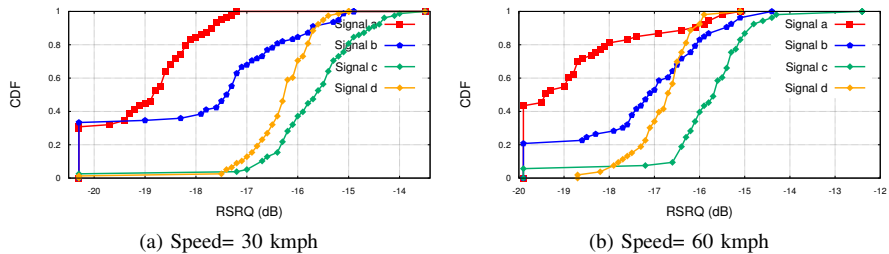}}
	\subfloat[Speed= 60 kmph]{\includegraphics[width=.33\linewidth]{figs/RSRQcdf120elv60}}
	\caption{A comparison of RSRQ CDF for signals with the highest RSRQ in different speeds and a fixed elevation of 120m.}
	\label{fig::RSRQcdfSpeed}
\end{figure*}

Finally, we show the empirical CDF of RSRQ of the serving cell's signal and its neighboring cells' signal for different elevations and different speeds in Fig. (\ref{fig::RSRQcdfElv} and \ref{fig::RSRQcdfSpeed}), respectively. We used the same naming of the signals in these figures as those of Fig. (\ref{fig::RSRPcdfElv} and \ref{fig::RSRPcdfSpeed}). Thus, the serving signals are the same as those of previous figures. 

%% file: conclusion.tex
\section{Conclusion}
\label{sec::conclusion}
Reusing the already deployed cellular network for low-altitude aerial user communication sounds promising, due to the cost efficiency, wide-coverage, high data transmission rate, and security of data transmission. However, the feasibility of such a communication, as well as its performance, need  comprehensive investigations. In this paper, we gathered data for an aerial LTE user in a rural area. We showed that for the low-altitude aerial users, the higher altitude leads to higher signal power, where the moderate altitude leads to slightly higher throughput. We did not experience any call drop during our tests which is a good sign for the feasibility of using the pre-deployed cellular network by the aerial users. The number of handover processes is very low at the highest altitude and the increment in the speed slightly decreases this number. While we tried to exhaustively study the reuse of the existed LTE network for aerial users, our study covered only a rural area. As future work, we aim at performing a similar study for suburban and urban areas. We can also compare the results for aerial users with that of terrestrial user via drive-test. 

\section{acknowledgment}
The authors acknowledge the financial and technical support provided by Infovista in using TEMS POCKET and TEMS Discovery in this project. The authors also acknowledge the cooperation of SEGA garden management.  